\documentclass[onecolumn,amsmath,amssymb,aps,superscriptaddress,longbibliography,10pt,floatfix]{revtex4-2}
\usepackage[T1]{fontenc}
\usepackage[utf8]{inputenc}
\usepackage{mathtools}
\usepackage{bm}
\usepackage{amsfonts}
\usepackage{mathrsfs}
\usepackage{braket}
\usepackage{dcolumn}
\usepackage{booktabs}
\usepackage{threeparttable}
\usepackage{makecell}
\usepackage{diagbox}
\usepackage{multirow}
\usepackage{tabularx}
\usepackage{graphicx}
\usepackage{chemfig}
\usepackage{chemformula}
\usepackage{subcaption}
\usepackage{color}
\usepackage{caption}
\usepackage{lipsum}
\usepackage{csquotes}
\usepackage{algorithm}
\usepackage{algorithmicx}
\usepackage{algpseudocode}
\usepackage{enumitem}
\usepackage{chngcntr}
\usepackage[colorlinks=true,citecolor=blue,linkcolor=blue,urlcolor=blue]{hyperref}
\usepackage{url}
\usepackage{orcidlink}
\usepackage{comment}
\usepackage{here}

\begin{document}
\title{Modeling phase separation in polymer-derived silicon carbonitride ceramics through extended machine learning molecular dynamics}

\author{Fabien Mortier}
\affiliation{Institut de Recherche sur les C\'eramiques (IRCER), UMR CNRS 7315, F-87068 Universit\'e de Limoges, Centre Européen de la Céramique, 12 rue Atlantis , Limoges, France.}
\author{Sylvian Cadars}
\affiliation{Institut de Recherche sur les C\'eramiques (IRCER), UMR CNRS 7315, F-87068 Universit\'e de Limoges, Centre Européen de la Céramique, 12 rue Atlantis , Limoges, France.}
\author{Olivier Masson}
\affiliation{Institut de Recherche sur les C\'eramiques (IRCER), UMR CNRS 7315, F-87068 Universit\'e de Limoges, Centre Européen de la Céramique, 12 rue Atlantis , Limoges, France.}
\author{Mauro Boero}
\affiliation{Universit\'e de Strasbourg, CNRS, Laboratoire ICube UMR 7357, F-67037 Strasbourg, France.}
\affiliation{ADYNMAT CNRS consortium, F-67034, Strasbourg, France.}
\author{Guido Ori}
\affiliation{Université de Strasbourg, CNRS, Institut de Physique et Chimie des Matériaux de Strasbourg, UMR 7504, F-67034 Strasbourg, France.}
\affiliation{ADYNMAT CNRS consortium, F-67034, Strasbourg, France.}
\author{Yun Wang}
\affiliation{Centre for Catalysis and Clean Energy, School of Environment and Science, Griffith University, Gold Coast, Australia}
\author{Samuel Bernard}
\affiliation{Institut de Recherche sur les C\'eramiques (IRCER), UMR CNRS 7315, F-87068 Universit\'e de Limoges, Centre Européen de la Céramique, 12 rue Atlantis , Limoges, France.}
\author{Assil Bouzid$^*$}
\email{assil.bouzid@cnrs.fr}
\affiliation{Institut de Recherche sur les C\'eramiques (IRCER), UMR CNRS 7315, F-87068 Universit\'e de Limoges, Centre Européen de la Céramique, 12 rue Atlantis , Limoges, France.}

\date{\today}

\begin{abstract}
\noindent\normalsize{Polymer-derived ceramics combine the thermal stability of ceramics with the versatile properties of carbon domains, but modeling their atomic-scale evolution during processing remains elusive due to the limitations of traditional computational methods. To address this issue, here we develop and apply a machine learning interatomic potential for silicon carbonitride-based (Si-C-N-H) systems, trained on a diversified database of over 9000 configurations - including amorphous models, high-temperature states, surfaces, and crystal structure predictions - to capture the full complexity of these materials. This potential enables large-scale molecular dynamics simulations of 8000-atom systems revealing the atomic-scale evolution of the polymer-derived ceramic during thermal treatment.
A key result of this work is the occurrence of a phase separation where carbon domains progressively nucleate from the amorphous SiCN matrix during thermal processing, forming distinct graphene-like sheets while preserving the integrity of the ceramic network. 
The resulting models reproduce the experimental atomic pair distribution functions with exceptional fidelity, validating our approach and providing microscopic explanations for the material unique combination of ceramic and graphitic properties. In this process, defective 5- and/or 7-member carbon rings, mediate the transformation to stable 6-member aromatic structures. These findings offer new atomic-scale insights into the thermal stability and structural transformation pathways of polymer-derived ceramics, while our methodology opens avenues for studying complex amorphous systems with first-principles accuracy at experimentally relevant scales.}
\end{abstract}

\maketitle

\section{Introduction}

Polymer-derived ceramics (PDCs) are advanced materials synthesized through pyrolysis of preceramic polymers such as polysilazanes, inorganic polysiloxanes or polycarbosilanes \cite{Colombo2010PDC, riedel2025preceramic}. This synthetic route, first pioneered by Fritz and Raabe in the 1950s and later expanded by Ainger, Herbert, and Chantrell in the 1960s, enables the production of ceramics with tailored compositions and microstructures at relatively low temperatures (800--1500$^\circ$C) compared to conventional ceramic manufacturing \cite{Fritz1956First_PDC, Ainger1960First_PDC, Chantrell1965First_PDC}. PDCs can adopt complex shapes and are compatible with several shaping techniques such as additive manufacturing, fiber spinning, or surface coating. In addition, thanks to their chemical and structural diversity, their thermal, electrical, and mechanical properties can be tuned \cite{Colombo2010Years, Riedel2006Synthesis, Chaudhary2022Additive, cheype2024straightforward}. Among the most studied PDC systems one can cite silicon carbide (SiC), silicon nitride (\ch{Si3N4}), silicon oxycarbide (SiOC), and silicon carbonitride (SiCN). These materials exhibit distinct microstructures and functional properties that are tightly linked to the synthesis procedure \cite{Eckel2016Additive, Barroso2019PDC}.\\

Among the PDCs, SiCN materials, when derived from polysilazane polymers pyrolysed at temperatures below 1400 $^\circ$C, feature a particularly interesting complex microstructure that arises from the nanoscale phase separation between an amorphous silicon-based network, typically composed of SiC$_x$N$_{4-x}$ tetrahedra, and a "free carbon" phase with a degree of organisation that depends on the pyrolysis temperature \cite{Wen2020Fate, Trassl2002Part-II,viard2017molecular}.
This phase separation is driven by the decomposition of organic groups during pyrolysis, leading to the precipitation of carbon-rich regions within the ceramic network. At temperatures below 1000$^\circ$C, the free carbon phase consists of small, disordered sp$^2$-hybridized C clusters, while at higher temperatures (above 1200 $^\circ$C), it progressively organizes into a turbostratic phase consisting of more-or-less extended stacked arrangements of potentially-defected graphitic layers, with various degrees of ordering up to graphite \cite{Trassl2002Part-II, Chen2014Quantitative}. 
Its presence is critical for boosting the overall electrical conductivity of the material, as sp$^2$-hybridized carbon domains exhibit metallic-like behavior, whereas the SiCN matrix remains semiconducting \cite{Trassl2003Electrical, Adigun2022Electrical}. Moreover, this free carbon phase acts as a diffusion barrier, and controlling its extent and degree of organization can prevent the crystallization of \ch{Si3N4} or SiC domains, thereby improving the thermal stability of the final material \cite{Kleebe2001Free_carbon, Mera2009Cristallization, klausmann2015synthesis,ding2024polymer}.\\

The functional properties of PDCs, particularly their electrical conductivity and catalytic activity, are closely tied to the organization of the free carbon phase. For instance, in metal-containing PDC such as SiCN-Ni or SiCN-Fe nanocomposites, the free carbon phase facilitates charge transfer between metal nanoparticles and the ceramic matrix, enhancing water splitting catalytic performance \cite{ferreira2023Low-Temperature, Ben_Miled2024Encapsulating}. 
However, if the carbon phase forms a dense or continuous network, it may also acts as a diffusion barrier, potentially hindering ionic transport \cite{Kleebe2001Free_carbon, Mera2009Cristallization, klausmann2015synthesis,ding2024polymer}. Thus, while electronic conductivity is improved, ionic diffusion could be compromised depending on the morphology of the carbon phase.
Within this context, the atomic-scale mechanisms governing the formation, growth, and interaction of this phase with the surrounding matrix are of primary importance, and yet remain poorly understood. Experimental techniques such as transmission electron microscopy (TEM), Raman spectroscopy, and solid-state NMR provide valuable insights into the structure of PDCs, and specifically the free carbon phase, but each has inherent limitations. TEM can detect graphitic domains but struggles to reveal the fine details of how these domains form at the atomic level or how they interact with the surrounding ceramic matrix \cite{Trassl2002Part-II}. Raman spectroscopy effectively identifies sp$^2$-hybridized carbon but cannot differentiate between small, isolated carbon clusters and larger, extended graphitic sheets \cite{Chen2014Quantitative}. Similarly, solid-state NMR offers detailed information about local C and Si atomic environments, but also lacks the longer-range information needed to fully map the phase separation within these complex materials \cite{Widgeon2012Carbodiimides}.
These limitations highlight the necessity of atomic-scale computational models to provide a complementary picture of the complex PDC composition-structure relationship escaping experimental probes.\\

Early computational studies of PDCs relied on empirical potentials, such as the Tersoff potential, or reverse Monte Carlo (RMC) methods. Dürr et al. \cite{Durr1998RMC} used RMC to model a \ch{Si24C43N33} composition, confirming phase separation between \ch{Si3N4}-like tetrahedra and amorphous carbon. Matsunaga et al. \cite{Matsunaga1999Tersoff} and Liao et al. \cite{Liao2012Tersoff} employed Tersoff potentials to study diffusion and mechanical properties, but these models lacked chemical reactivity and turned out to be insufficient to capture mixed SiC$_x$N$_{4-x}$ environments. As shown by Kroll, first-principles molecular dynamics (FPMD) on \ch{Si40C40N40} systems demonstrated the occurrence of phase separation and the formation of carbon clusters, but the small system sizes (120 atoms) and short simulation times (40 ps) limited insights into the long-term structural evolution \cite{Kroll2005Modelling}. In another work, Amkreutz et al. \cite{Amkreutz2002Understanding} used density functional-based tight binding (DFTB) to study the phase separation in \ch{Si37C32N31} systems, proposing that the free carbon phase could exist as either small clusters or extended layered structures. While insightful, these studies left several open questions about the dynamical evolution of the free carbon phase.\\

In a recent FPMD-based study \cite{mortierfpmd2025} we addressed some of these limitations by focusing on the \ch{Si32C25N24H19} PDC derived from \ch{-[Si(CH3)(CH=CH2)-NH-]_{0.2}-[Si(H)(CH3)-NH-]_{0.8} -} polyvinylsilazane. Using the ceramicNetworkBuilder algorithm (available in the Python Atomic Modeling Analyzes library \cite{PyAMA}), we generated several initial configurations differing in the organization and networking of the free carbon phase, including clusters, chains, and pre-inserted graphitic layers. After thermal annealing, the models were validated against experimental pair distribution functions (PDFs), revealing that models where carbon is essentially made of poorly spatially correlated layers, rather than extended and stacked layers, best match the experimental data. While this work provided an atomic-scale validation of plausible free carbon phase structures and atomic local environment within PDCs, it is still constrained by the small system sizes (400 atoms) and short timescales (hundreds of picoseconds) accessible to FPMD. As such, it did not capture the medium/long-range dynamics of carbon domain growth or the full complexity of phase separation over experimentally relevant time- and/or size-scales.\\

The breakthrough of machine-learning interatomic potentials (MLIPs) has revolutionized modeling of complex materials by combining accuracy of ab initio methods with the efficiency of classical force fields. As far as PDC materials are concerned, Kroll et al. \cite{Kroll2024, Haseen2025} recently introduced machine-learning interatomic potentials for Si-C-N-H and Si-C-O-H systems based on moment tensor potentials (MTP). In these studies, the reported final training sets comprise 3417 structures for Si-C-O-H \cite{Kroll2024} and 2104 structures for Si-C-N-H \cite{Haseen2025}. While both articles describe the principal sources of configurations (e.g., molecules, polymers, crystalline and amorphous structures, and high-temperature trajectories), the detailed sampling distribution across compositions, thermodynamic conditions, and reactive motifs is not fully specified. In addition, the need to increase the MTP order level for chemically complex and reactive environments, particularly in Si-C-N-H systems \cite{Haseen2025}, highlights an important accuracy–efficiency trade-off.\\

Despite these limitations, Kroll et al. demonstrated that MTP-based MLIPs can reproduce key near-equilibrium properties and enabled the simulation of polymer-to-ceramic conversion at size and time scales inaccessible to ab initio methods. For Si-C-O-H, the MTP-MLIP reproduced vibrational signatures and captured the main features of polysiloxane pyrolysis, including the emergence of mixed Si-centered tetrahedra and the precipitation of graphitic carbon in C-rich compositions, with simulations reaching million-atom models over nanoseconds \cite{Kroll2024}. For Si-C-N-H systems, the MTP-MLIP enabled pyrolysis simulations of polysilazanes to amorphous SiCNH with plausible carbon-segregation morphologies and supported kinetic analysis from simulated thermogravimetric data \cite{Haseen2025}.
Nevertheless, as these potentials were primarily developed to enable reactive simulations of polymer pyrolysis, the nucleation and growth of the free-carbon phase, as well as its interaction with the surrounding SiCO/SiCN matrix, were not systematically investigated as central objectives of those works. Furthermore, a direct quantitative comparison of the predicted carbon-phase morphology within the SiCN matrix with experiment remains limited. This makes it difficult to assess the reliability of the MLIPs specifically for describing the final structure of the phase separated ceramic.\\

In the present study, we overcome these limitations by developing and applying an accurate MLIP for Si-C-N-H systems, trained on a large database containing over 9000 configurations featuring a high degree of structural and compositional diversity to ensure model robustness. This database includes various amorphous bulk Si-C-N-H models, high-temperature overheated states, surfaces and interface models to describe free carbon domain boundaries, crystal structure prediction models to sample rare but potentially critical bonding environments, and amorphous carbon and silicon structures to ensure transferability across extreme compositions. Our MLIP, built using the MACE framework \cite{Batatia2023MACE, Batatia2025MACE2}, achieves a remarkable accuracy on an unseen test database, with root mean square errors of 12.4 meV/atom for energies and 149 meV/{\AA} for forces. This enables large-scale molecular dynamics simulations of 8000-atom systems over nanosecond timescales, providing new insights into the atomic-scale mechanisms driving carbon domain nucleation and growth in amorphous Si-C-N-H systems. \\

Overall, we show that by relying on a structurally diversified database together with equivariant machine-learning interatomic potentials, we establish a robust and generalizable framework for the study of complex phenomena in amorphous ceramics. 

\section{Methods}

\subsection{Database construction}

The MLIP training database consists of seven distinct classes of configurations, carefully selected to represent the structural diversity of Si-C-N-H systems (Table \ref{tab:database}). It contains over 9000 configurations spanning a wide range of compositions, densities, and structural motifs and networking.

\begin{table}[!ht]
\caption{Composition of the training database for the Si-C-N-H MLIP.}
\label{tab:database}
\centering
\small
\begin{tabular}{lccc}
\toprule
Class & Composition & Atom Count & Number of Configurations \\
\midrule
SiCNH-periodic & SiCNH & 200 and 400 & 4033  \\
SiCNH-periodic-4000K & SiCNH & 400 & 1497  \\
SiCNH-surface & SiCNH & 200 and 400 & 2000  \\
Amorphous-Carbon-200 & C & 216 & 249  \\
Amorphous-Carbon-250 & C & 216 & 249  \\
Amorphous-Silicon & Si & 512 & 250  \\
Crystal-Structure-Prediction & SiCNH & 46 to 100 & 1165  \\
\midrule
Total & & & 9443 \\
\bottomrule
\end{tabular}
\end{table}

To ensure the consistency of the database, the energies and forces of all the collected configurations were recalculated by resorting to the CP2K code \cite{Kuhne2020CP2K} using the same density functional theory (DFT) parameters with the PBE functional \cite{PBE1996Generalized}, GTH pseudopotentials \cite{GTH1996Separable}, double-$\zeta$ polarized basis sets \cite{VandeVondele2005Quickstep}, a plane-wave energy cutoff of 1000 Ry, and semi-empirical Grimme's D3 dispersion corrections \cite{Grimme2010DFT-D}. For smaller systems (46-100 atoms) obtained from crystal structure prediction (using a different DFT code, as will be detailed below), $\Gamma$-point calculations are not sufficient. These systems were therefore subjected to a single-point energy calculation in CP2K, using the same parameters as the FPMD described above, with the addition of k-point meshes constructed automatically to ensure a maximum spacing of 0.3 {\AA}$^{-1}$ between k-points in the reciprocal space.

\subsubsection{Configurations derived from first-principles molecular dynamics}

The SiCNH-periodic class contains configurations extracted from FPMD trajectories of periodic models with composition \ch{Si32C25N24H19}. This class includes configurations from seven models presented in our previous work \cite{mortierfpmd2025} in addition to three new models produced with the purpose of increasing the amorphous configuration diversity. Details of the three new models are provided in the supplementary materials paragraph S1. 125 configurations were sampled from each temperature plateau of the specific thermal cycles applied to every models (the temperature ranging from 300 K to 1800 K). It results in a wide database of more than 7000 configurations, from which approximately 4000 are selected to feed the training, validation and test sets employed to develop the MLIP.\\

The SiCNH-periodic-4000K class contains high-temperature configurations (up to T = 4000 K) generated from model SiCNH400-A of Ref. \cite{mortierfpmd2025} to sample high-energy states. This class is particularly important for ensuring that the MLIP can accurately describe configurations far from equilibrium, which may be encountered during phase separation or thermal processing. The high-temperature configurations were generated using the thermal cycle presented in Figure S1 in the supplementary materials, resulting in 1497 configurations.\\

The SiCNH-surface class contains configurations from five surface models, with and without exposed carbon layers/chains, produced using the procedure described in the supplementary materials paragraph S3. Similar to the periodic models, 125 configurations were sampled from each temperature plateau (T = 1300, 900, 600 and 300 K), resulting in 2500 configurations. The inclusion of these surface configurations aims at capturing the structural motifs that may appear at interfaces or in the vicinity of free carbon domains.

\subsubsection{Amorphous Carbon and Silicon configurations}

The Amorphous-Carbon-200 and Amorphous-Carbon-250 classes incorporate amorphous carbon structures at densities of 2.00 and 2.50 g/cm$^{3}$, respectively. Initial configurations were obtained from the work of Deringer and Csányi \cite{Deringer2017Carbon} where melt-quenched models were produced at different densities, resulting in models with carbon sp, sp$^2$, and sp$^3$ hybridizations. Each initial amorphous carbon configuration was subjected to a short thermal cycle of about 20 ps at T = 300 K, from which 249 configurations were extracted from each model. \\

Similarly, for the Amorphous-Silicon class, we collect initial configuration  from the work of Deringer et al. \cite{Deringer2018Silicon}. Specifically, we consider the model containing 512 silicon atoms obtained from a melt-quenching simulation with a quenching rate of 10$^{11}$ K/s. The initial configuration was subjected to a short thermal cycle of 20 ps at T = 300K, from which 250 configurations were included into our database. \\

The inclusion of these configurations is of paramount importance for describing the carbon or silicon-rich environments that may be encountered in certain regions of the SiCNH matrix and particularly the free carbon domains that may nucleate and grow during a phase separation.

\subsubsection{Crystal Structure Predictions}

The Crystal-Structure-Prediction class contains 1165 configurations generated using the USPEX program \cite{Oganov2006USPEX1, Oganov2011USPEX2, Lyakhov2013USPEX3} for various Si-C-N-H compositions and densities. USPEX is a crystal structure prediction (CSP) program employing evolutionary algorithm to generate, generation after generation, a diverse set of candidate structures for a given composition. While the purpose of this approach is generally to efficiently explore the potential energy surface and determine the most stable crystalline structure, here we exploit, instead, the structural diversity resulting from this exploration. Each structure generated with the evolutionary algorithm is relaxed with DFT, using the VASP code \cite{kresse1993ab,kresse1996efficient}, within 3 consecutive steps of increasing numerical accuracy, followed by a high-accuracy single-point energy calculation (see the supplementary materials paragraph S4 for details).

\begin{table}[!ht]
\caption{Configurations generated by crystal structure prediction (Crystal-Structure-Prediction class).}
\label{tab:uspex_compositions}
\centering
\small
\begin{tabular}{lccc}
\toprule
\multirow{2}{*}{Name} & \multirow{2}{*}{Composition} & \multicolumn{2}{c}{Number of Configurations} \\
{} & {} & Optimized density & Low density \\
\midrule
Ref-compo & \ch{Si16C12N12H10} & 110 & 0 \\
Ref-compo-low-density & \ch{Si32C24N24H20} & 0 & 144 \\
C-rich & \ch{Si12C20N10H8} & 120 & 0 \\
C-poor & \ch{Si16C8N12H10} & 115 & 0 \\
N-rich & \ch{Si12C10N20H8} & 114 & 0 \\
N-poor & \ch{Si18C14N6H12} & 104 & 0 \\
H-rich & \ch{Si12C9N9H20} & 139 & 131 \\
C-H-rich & \ch{Si8C16N8H18} & 104 & 84 \\
\bottomrule
\end{tabular}
\end{table}

The CSP configurations were generated for seven distinct compositions representing variations around the target material composition, with some compositions independently explored at optimized and (fixed) low density (see Table \ref{tab:uspex_compositions}). The low-density configurations were generated by fixing the volume to 1.25 times the average volume of the 10 best structures from the variable-density calculations conducted for the same composition. For each composition and density, 1000 to 2000 structures were generated, and 100 to 200 were selected to build the database (see supplementary materials paragraph S4). \\

\begin{figure}[!ht]
\centering
\includegraphics[width=0.6\linewidth]{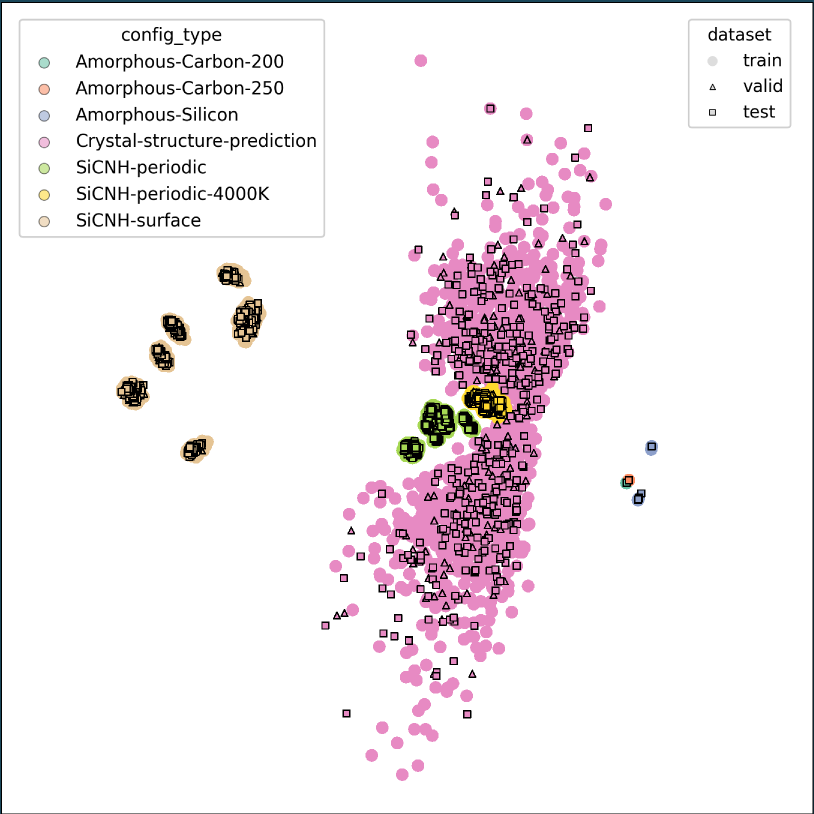}
\caption{Multidimensional scaling 2D projection of the training database using Valle-Oganov fingerprints \cite{Valle2010Fingerprint}, demonstrating the structural diversity captured by the database. Each point represents a configuration, with colors indicating different classes of structures.}
\label{fig:database_map}
\end{figure}

The inclusion of these crystalline configurations is crucial for ensuring that the MLIP can accurately describe structures that may be far from the typical amorphous configurations encountered in FPMD simulations. This diversity is particularly visible in the multidimensional scaling \cite{kruskal1964nonmetric,pedregosa2011scikit} 2D projection of the training database using Valle-Organov fingerprints \cite{Valle2010Fingerprint} as given in Figure \ref{fig:database_map}. It is very important for capturing the complex energy landscape of Si-C-N-H systems and ensuring the transferability of the potential across different structural motifs.

\subsection{MLIP training}

The interatomic potential was developed using the MACE framework \cite{Batatia2023MACE, Batatia2025MACE2} (version 0.3.9), which implements a message-passing neural network architecture that has shown exceptional performance for a variety of materials systems \cite{kovacs2023evaluation, barrett2025transferable, roy2026comparison}. The architecture consists of 2 message-passing layers with 128 scalar ($L=0$) and 128 vector ($L=1$) components, providing a balance between computational efficiency and descriptive power. The correlation order $\nu$ was set to 3, allowing the potential to capture 4-body interactions (one central atom and three neighbors). Such level of correlation aims at accurately describing the complex bonding environments in Si-C-N-H systems, where the coordination of silicon and carbon atoms can vary significantly depending on the local chemical environment. The atomic cluster expansion descriptor is used to represent atomic environments and uses spherical harmonics up to degree $l_{max}=3$ and 8 radial basis functions derived from 5 Bessel functions. The cutoff radius was set to 5.5 {\AA}, which corresponds to an effective receptive field of 11 {\AA}, thereby providing a good balance between capturing long-range interactions and maintaining computational efficiency. The model training was performed using Adam optimizer \cite{Diederik2017Adam} with a batch size of 10 configurations.\\

Given the complexity of the explored energy landscape, the training was conducted gradually in five subsequent steps, wherein the training database was progressively expanded, as shown in Table \ref{tab:training_phases}. This gradual procedure allows the potential to first learn the basic features of the energy landscape before being exposed to more specific or complex configurations.

\begin{table}[!ht]
\caption{Training steps for the Si-C-N-H MLIP.}
\label{tab:training_phases}
\centering
\small
\begin{tabular}{lccccc}
\toprule
Class & DB-01 & DB-02 & DB-03 & DB-04 & DB-05 \\
\midrule
SiCNH-periodic       & X & X & X & X & X \\
SiCNH-surface        &   &   &   &   & X \\
SiCNH-periodic-4000K &   &   &   & X & X \\
Amorphous-Carbon-200 &   &   & X & X & X \\
Amorphous-Carbon-250 & X & X & X & X & X \\
Amorphous-Silicon    &   & X & X & X & X \\
Crystal-Structure-Prediction& X & X & X & X & X \\
\bottomrule
\end{tabular}
\end{table}

Each training step consisted of two stages: an initial stage focusing on forces optimization and a subsequent stage focusing on energy refinement. In practice, this is governed by the modification of the learning rate and the weights of the loss function $\mathcal{L}$ given in Equation \ref{Eq_loss_MACE}. 

\begin{equation}\label{Eq_loss_MACE}
    \mathcal{L} = \frac{\lambda_{E}}{B}\sum_{b=1}^{B}\left(\frac{E_{b}-\hat{E}_{b}}{N_{b}}\right)^{2} + \frac{\lambda_{F}}{3B}\sum_{b=1}^{B}\sum_{i_{b},\alpha=1}^{N_{b},3}\left(-\frac{\partial{E_{b}}}{\partial{}r_{i_{b},\alpha}}-\hat{F}_{i_{b},\alpha}\right)^{2}
\end{equation}

Here, $\mathcal{L}$ corresponds to a weighted sum of the mean square errors of the total energy and forces components over the atomic configurations $b$ containing $N_{b}$ atoms in a batch of size $B$. $E_{b}$ depicts the energy predicted by the model while $\hat{E}_{b}$ and $\hat{F}_{i_{b},\alpha}$ correspond to the DFT configuration energy and forces components of the atom $i$ of configuration $b$ along the $\alpha$ direction, respectively.
During the force optimization stage, the learning rate was set to $lr$ = 0.01, with loss weights of $\lambda_F=100.0$ for forces and $\lambda_E=1.0$ for energies and during the energy refinement stage, the learning rate was reduced to $lr$ = 0.001, with loss weights of $\lambda_F=100.0$ for forces and $\lambda_E=1000.0$ for energies. This two-stage approach allows the potential to first adjust the model hyperparameters with less tight constraints before refining them to describe the subtle energy differences between configurations.\\ 

Furthermore, at each training step, each new database introduced is first split into 80\% training-validation and 20\% test sets. This ensures that 20\% of both the whole dataset and of each individual database (SiCNH-periodic, SiCNH-surface, CSP, etc.) remains totally unseen by the training process. Subsequently, the training-validation set is split into 90\% training and 10\% validation sets. The training set is used for training the model, while the validation set was used to monitor the validation loss during the training and prevent overfitting. The final model performance is evaluated on the totally unseen test set.\\ 

\begin{figure}[!ht]
\centering
\includegraphics[width=0.8\linewidth]{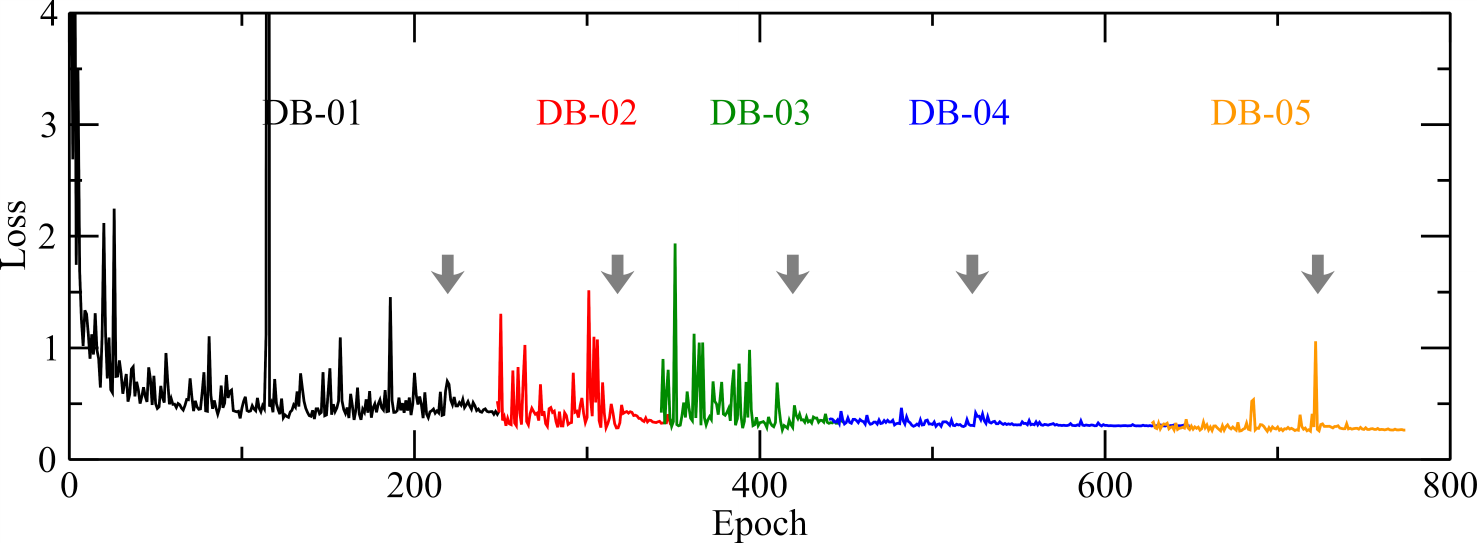}
\caption{Evolution of the loss on the validation set during the training process. The figure shows the two stages of each training phase, with the initial focus on force optimization followed by energy refinement. Arrows indicate switches from the force-optimization to the energy-optimization stage.}
\label{fig:training_convergence}
\end{figure}

The training process was conducted over 775 epochs, with the final MLIP demonstrating exceptional accuracy and transferability across the entire database. Figure \ref{fig:training_convergence} shows the evolution of the loss function on the validation set during the training process where arrows indicate the force to energy optimization stage changes for every training step. Figure S4 in the supplementary materials displays the evolution of the associated root mean square errors (RMSE) for energies and forces. In both training stages, force errors decrease at first, followed by a more gradual improvement in energy prediction accuracy.

\subsection{MLIP validation}

\begin{figure}[!ht]
\centering
\includegraphics[width=0.45\linewidth]{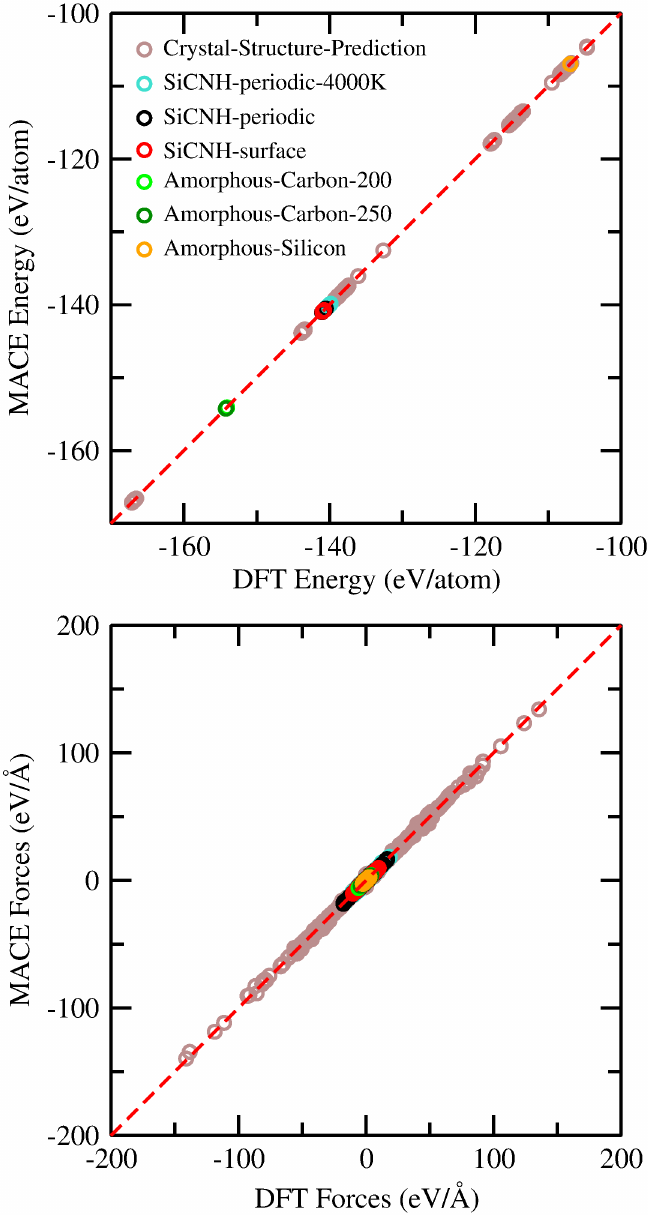}
\caption{Comparison of MLIP-predicted energies and forces with DFT reference values for the test set.}
\label{fig:validation}
\end{figure}

Figure \ref{fig:validation} shows the comparison between final MLIP predictions and DFT reference values for energies and forces across all configuration classes in the (previously unseen) test set. The performance of the potential is quantified using root mean square errors (RMSE) and relative RMSE (RRMSE) for each configuration class (see Table \ref{tab:rmse}). The overall RMSE of the model are 12.4 meV/atom for energies and 149 meV/{\AA} for forces, with a RRMSE of 10.2\% for forces. 
By looking at configuration classes individually, we observe that RMSE on energies fall below 10 meV/atom and RMSE on forces are of the order of 100 meV/\AA{} for SiCNH-periodic, SiCNH-surface, Amorphous-Carbon-200, Amorphous-Carbon-250 and Amorphous-Silicon. For SiCNH-periodic-4000K, while RMSE on energy is also below 10 meV/atom, the value on forces reaches 218 meV/\AA{}. This rise in forces RMSE comes from the higher energy of the configurations produced at high temperature (> 2000 $^\circ$C) included in this class. This higher absolute RMSE is nevertheless not followed by a significant increase of the RRMSE which is of the same order than the RRMSE of the other classes of configuration (between 8 and 13\%). 
The absolute RMSE on both energies and forces are relatively higher for the Crystal-Structure-Prediction configurations than for other classes (due to the higher energies and forces of these configurations). However, its RRMSE on forces is only of 9.2\%. These performances are particularly noteworthy because crystalline configurations of the Crystal-Structure-Prediction class contain many high-energy atomic environment far from the typical amorphous structures encountered in FPMD simulations. They highlight the accuracy and transferability of the potential, making it suitable for large-scale molecular dynamics simulations of Si-C-N-H systems. In the following sections, we apply this potential to study the atomic-scale mechanisms governing carbon nucleation and growth in Si-C-N-H PDCs.

\begin{table}[!ht]
\caption{RMSE and RRMSE values for the MLIP on the test set, compared to DFT reference calculations.}
\label{tab:rmse}
\centering
\small
\begin{tabular}{lccc}
\toprule
Class & RMSE Energy & RMSE Forces & RRMSE Forces \\
& (meV/atom) & (meV/{\AA}) & (\%) \\
\midrule
SiCNH-periodic & 7.1 & 128 & 9.5 \\
SiCNH-surface & 3.0 & 104 & 8.5 \\
SiCNH-periodic-4000K & 3.3 & 218 & 12.6 \\
Amorphous-Carbon-200 & 7.9 & 110 & 12.0 \\
Amorphous-Carbon-250 & 7.8 & 103 & 10.9 \\
Amorphous-Silicon & 1.1 & 51.3 & 9.1 \\
Crystal-Structure-Prediction & 31.8 & 373 & 9.2 \\
\midrule
Total & 12.4 & 149 & 10.2 \\
\bottomrule
\end{tabular}
\end{table}

\subsection{Structural models generation}

Molecular dynamics simulations were performed using the LAMMPS software \cite{Thompson2022LAMMPS}. LAMMPS was configured with the corresponding mace pair style and calculations were performed on 1 GPU, enabling simulation of  thermal cycles lasting for a few nanoseconds on systems of 8000 atoms. To fully exploit modern GPU architectures, we compiled LAMMPS with Kokkos+CUDA \cite{edwards2014kokkos} and message parallel interface, explicitly targeting the NVIDIA H100 architecture. All simulations were conducted in the NVT ensemble using periodic boundary conditions, a time step of 0.5 fs and thermal control through a Nosé-Hoover thermostat \cite{Nose1984Canonical,Nose1984Unified,Hoover1985Canonical}, ensuring an optimal balance between computational efficiency and accurate integration of the equations of motion.\\

Four models of composition \ch{Si32C25N24H19} were generated based on three initial configurations produced at the experimental density of 2.17 g/cm$^{3}$, corresponding to a cubic cell of side length of around 4.5 nm, using different approaches to investigate the influence of initial atomic arrangement on the final structure after thermal annealing. This composition has been extensively studied in our previous FPMD work, using models up to 400 atoms \cite{mortierfpmd2025}, which provided the primary brick of our database.\\ 

The initial configuration of the \textbf{SiCNH8000-Random} model corresponds to a random distribution of atoms generated using the Packmol software \cite{Martinez2003Packing, Martinez2009Packmol}. This approach provides a completely unbiased starting point for the simulation, with no a-priori knowledge about the network connectivity beyond the local coordination environment of the considered species. It also allows to assess the ability of the MLIP to describe highly random structures.\\ 

The \textbf{SiCNH8000-CNB} configuration was generated using the ceramicNetworkBuilder program  with parameters that were explored and adjusted in our previous work \cite{mortierfpmd2025}. This approach leads to an amorphous network based on specified pair-bonding and coordination probabilities, providing a more realistic starting point than the random configuration. We here used parameters that favor the formation of small carbon motifs (chains, clusters) in the initial model. The target bonding and coordination number probabilities employed to generate the initial configuration of the SiCNH8000-CNB model are given in Tables S2 and S3 in the supplementary materials.\\

For the \textbf{SiCNH8000-CL} configuration, prior to generating the amorphous network, 46 carbon sheets (27 with 3 rings and 19 with 4 rings) were inserted into the simulation cell, representing 32.5\% of the carbon atoms in the system. Next, the ceramicNetworkBuilder program was used to complete the amorphous network (see parameters in Tables S2 and S3 in the supplementary materials). Carbon atoms outside of the pre-formed carbon layers were forced to bond only to silicon in order to diversify the population of SiC$_x$N$_{4-x}$ tetrahedra in the amorphous network of the ceramic, by creating more Si-C$_2$N$_2$ or Si-C$_3$N units as evidenced by NMR analysis \cite{Seitz1996Bonds}.
This approach provides a starting point with pre-formed carbon domains, allowing us to study their evolution during thermal cycling.\\

Figure S5 in supplementary materials shows the initial configurations of the three models. The distribution of carbon atoms varies significantly between the models, from isolated atoms and small clusters in the SiCNH8000-Random model to extended pre-formed rings and clusters in the SiCNH8000-CL model.\\

\begin{figure}[!ht]
\centering
\includegraphics[width=0.6\linewidth]{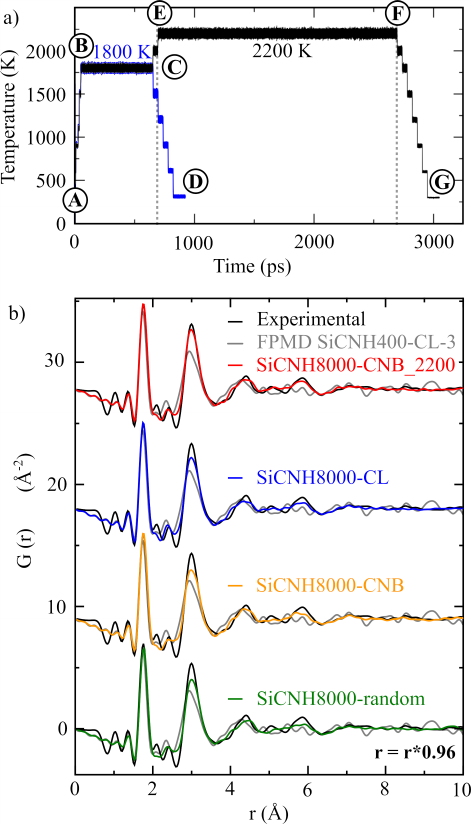}
\caption{a) Thermal cycles applied to the 8000-atom models using the MLIP. The extended high-temperature thermal cycle shown in black was used to obtain the SiCNH8000-CNB\_2200 model, while other models were obtained with the shorter thermal cycle shown in blue. b) Total pair distribution functions of the 8000-atom models compared to experimental data and FPMD SiCNH400-CL-3 from Ref. \cite{mortierfpmd2025}. Simulated data use $r$ values scaled by 0.96 for optimal alignment with experimental peaks (see text).}
\label{fig:total_pdf}
\end{figure}

The three starting configurations were subjected to the same thermal cycle using MLIP molecular dynamics (Figure \ref{fig:total_pdf}a). It consists of subsequent temperature plateaus separated by 2 ps transitions to ensure smooth temperature changes. The cycle started with a first stabilization phase of 3 ps at T = 300 K, followed by a heating phase consisting of 2 ps at T = 600 K, 18 ps at T = 900 K, 8 ps at T = 1100 K, 10 ps at T = 1500 K, a long plateau of 600 ps at the maximum temperature of T = 1800 K, a cooling phase consisting of 40 ps at T = 1500 K, 40 ps at T = 1200 K, 40 ps at T = 900 K and 40 ps at T = 600 K, and finally 100 ps at T = 300 K. This thermal cycle is designed to study the structural evolution of the SiCNH material during annealing phases within reasonably affordable simulation times.\\ 

To favor possible growth of the carbon domains, we extended the thermal cycle  for the SiCNH8000-CNB model, starting from the end of the T = 1800 K plateau to reach a temperature of T = 2200 K (via an intermediate plateau at T = 2000 K), and maintained this elevated temperature for 2 ns, after which the system was brought down to room temperature. The fourth model obtained at the end of this extended thermal cycle is henceforth referred to as \textbf{SiCNH8000-CNB\_2200}.

\subsection{Validation of the Si-C-N-H models}

Figure \ref{fig:total_pdf}b presents a comparative analysis of the total PDFs of our four 8000-atom models and the experimental reference data \cite{mortierfpmd2025}. For the sake of comparison, we also show the best model obtained from previous FPMD simulations on a 400 atoms cell from Ref. \cite{mortierfpmd2025}. 

As discussed in Ref. \cite{mortierfpmd2025}, the FPMD Si-C-N-H amorphous models PDFs show a better agreement with the experimental PDF (especially for peaks positions) when applying an $r$ scaling factor slightly lower than 1. 
This scaling factor was explained mainly by the adopted DFT setup that tends to slightly overestimate atomic bond lengths. The resulting r-scaled model density was 11\% larger than the experimental one, which was attributed to an underestimation of the measured density (by helium pycnometry) due to the presence of a large fraction of closed porosity in the sample \cite{mortierfpmd2025}. Given that our MLIP was trained on the FPMD data, the same trends are also observed in the MLIP generated models studied in this work. Indeed, the best agreements with experimental data are obtained by applying similar r-scaling factors of approximately 0.96 (see Figure \ref{fig:total_pdf}b) leading to a models density 13\% larger compared to the experimental value.\\

The PDFs simulated from all four MLIP-MD models reveal a considerably increased agreement with experimental data, compared with the 400-atom FPMD model, regardless of their different initial configurations.
Focusing on peak intensities, the two experimental PDF peaks centered at 1.37 {\AA} and 1.76 {\AA} are well reproduced by all our MLIP-MD models. These peaks correspond respectively to C-C bonds and Si-C and Si-N bonds, as previously identified \cite{mortierfpmd2025,Schempp1998Study}. In addition, the C-(C)-C peak at 2.42 {\AA}, which in our previous FPMD models (particularly those with large stacked carbon sheets) was often overestimated, now shows a better match with the experimental PDFs, suggesting an improved description of the atomic-scale structure as model size and simulation times are increased. 
The most striking difference between the PDFs of the FPMD model and of the 8000-atom MLIP-MD models is observed for the peak centered at around 3.00 {\AA}, which features increased intensity and slightly reduced width in the latter, in better agreement with experiment, particularly for the SiCNH8000-CNB\_2200 model. This peak encompasses multiple correlations, including N-(Si)-N, N-(Si)-C, Si-(N)-Si, Si-(C)-Si, and C-(C-C)-C. The intensity difference between calculated and experimental PDFs was previously attributed (based on FPMD models) to the presence of Si-Si bonds or -Si-X-Si-Y- 4-member rings (X, Y = N or C) disrupting the organization of the SiCN network. The increased intensity of this peak at 3.00 {\AA} in the large MLIP-MD models therefore suggests a slight improvement in the description of the network when using the MLIP potential, which can be attributed to the large time and size-scales of these simulations allowing the system to achieve a better relaxation. Finally, while the peaks located beyond 4 {\AA} remain broad and of low intensity, indicating a limited order at these longer correlation lengths, they are better reproduced by the large models than by previous FPMD models.\\ 

Overall, the PDFs of our 8000-atom models exhibit a notably better agreement with the experimental reference, compared to models obtained from FPMD simulations. Focusing on the MLIP models, SiCNH8000-CL and particularly SiCNH8000-CNB\_2200 show an overall better agreement with the experimental PDF than SiCNH8000-Random and SiCNH8000-CNB models. This is particularly visible on the large $r$ values.\\

In order to quantitatively compare the experimental and the modeling results, we minimized the RMSE between experimental and simulated data by optimising the $r$ scaling factor. In addition, we also optimized the amplitude of the calculated PDFs, as imperfect data normalization procedure might affect the experimental PDF intensity scaling. The obtained optimized PDFs are shown in Figure S6 in the supplementary materials and the corresponding RMSE results are summarized in Table \ref{tab:pdf_rmsd}. This analysis confirm the superiority of  SiCNH8000-CL and particularly SiCNH8000-CNB\_2200 models in reproducing the experimental PDFs. 

\begin{table}[!ht]
\caption{Results of optimisation of simulated PDF amplitude and $r$ scaling for best match with experimental data.}
\label{tab:pdf_rmsd}
\centering
\small
\begin{tabular}{lcccccc}
\toprule
Model & Amplitude scaling & $r$ scaling & RMSE \\
\midrule
SiCNH400-CL-3 (FPMD, \cite{mortierfpmd2025}) & 1.313 & 0.977 & 0.521 \\   
SiCNH8000-Random     & 1.290 & 0.958 & 0.311 \\   
SiCNH8000-CNB        & 1.233 & 0.961 & 0.315 \\   
SiCNH8000-CL         & 1.237 & 0.962 & 0.289 \\   
SiCNH8000-CNB\_2200   & 1.147 & 0.961 & 0.260 \\  
\bottomrule
\end{tabular}
\end{table}

\section{Results and discussion}
\subsection{Structural characterization of the amorphous models}
\subsubsection{Partial pair distribution functions}

\begin{figure}[!ht]
\centering
\includegraphics[width=0.7\linewidth]{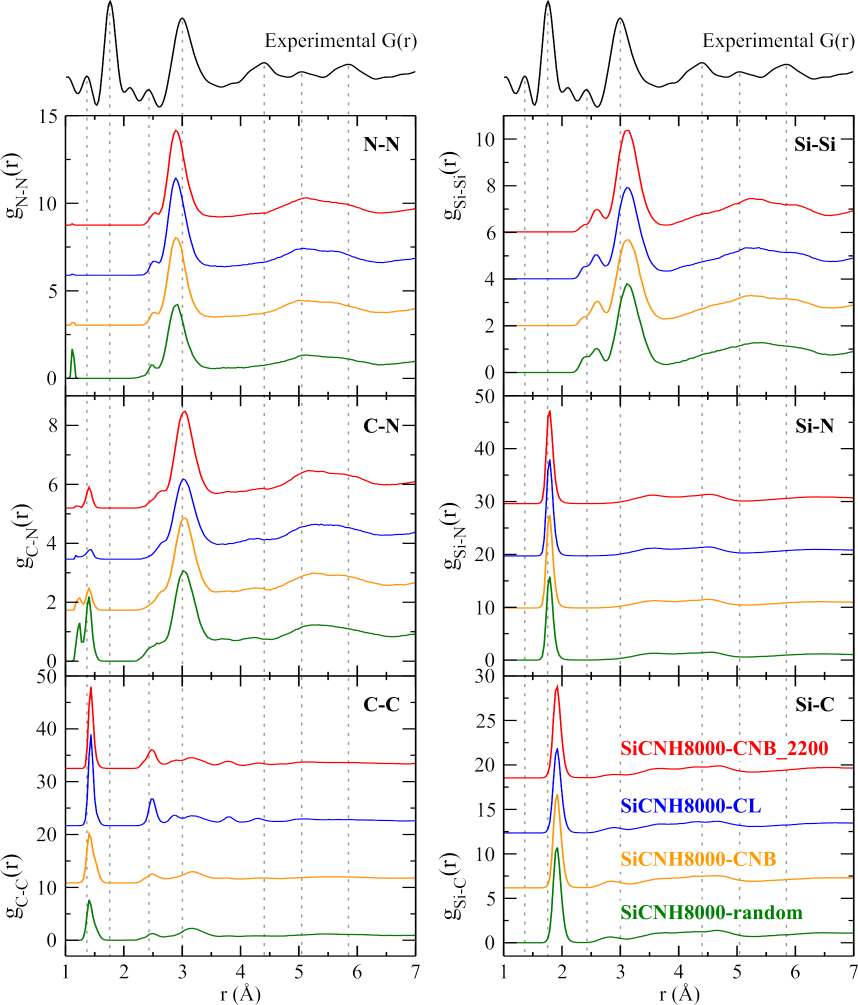}
\caption{Partial pair distribution functions for the 8000-atom models.}
\label{fig:partial_pdfs}
\end{figure}

The models being validated, we now focus on their structural characterization. Figure \ref{fig:partial_pdfs} presents the partial PDFs for the four 8000-atom models excluding hydrogen-containing pairs, which provide only very small contributions to the total PDF. Before examining the individual partial PDFs, we first point that most major partials peaks not only have considerably smoother and more regular shapes than the corresponding peaks in the previous FPMD partials in Ref. \cite{mortierfpmd2025} (which is primarily due to better statistics in 8000-atom models), but they are also significantly narrower. This suggests that many irregular environments present in FPMD models (obtained after a few tens of ps of annealing) are eliminated by the ns timescales accessible to MLIP simulations.

First, the $\textrm{g}_{\textrm{C-C}}(r)$ functions highlight the increasing organization of carbon domains when moving from SiCNH8000-Random to  SiCNH8000-CNB, SiCNH8000-CL and finally to SiCNH8000-CNB\_2200 model.
The first peak centered at 1.42 {\AA} is more intense for SiCNH8000-CL and SiCNH8000-CNB\_2200 models compared to the other two, indicating a greater quantity of C-C bonds in these models. The sharpening of this peak also suggests a tendency for carbon atoms to adopt an sp$^2$ configuration. Interestingly, while SiCNH8000-Random and SiCNH8000-CNB models show broad C-C correlations beyond 2 {\AA}, SiCNH8000-CL and SiCNH8000-CNB\_2200 models exhibits well defined peaks at distances compatible with those found in typical C rings ($\approx$ 2.47 and 2.9 {\AA}).\\ 

The $\textrm{g}_{\textrm{N-N}}(r)$ functions of the MLIP-generated models display a peak at approximately 1.15 {\AA}, indicating the presence of nitrogen molecules in these models. The quantities of these N$_2$ molecules, formed during the early stages of the molecular dynamics thermal cycle due to defects in the initial configurations, are 28, 2, 1, and 1 in the SiCNH8000-Random, SiCNH8000-CNB, SiCNH8000-CL and SiCNH8000-CNB\_2200 models, respectively. The random generation procedure has a non-negligible probability to form these (undesired) species, whereas the ceramicNetworkBuilder approach (with $p(N$-$N)=0$) considerably reduces this probability by preventing the formation of N-N bonds in initial models. However, the ceramicNetworkBuilder-based initial configurations still contain defects such as under-coordinated nitrogen atoms that, when sufficiently close to one another, can stabilize by forming N$_2$ molecules. 
The small peak around 2.53 {\AA} is due to the presence of four-atom -Si-N-Si-N- cycles in the models and the most intense N-N peak at 2.92 {\AA} represents the N-(Si)-N distances of the SiCN network.  \\

Focusing on the $\textrm{g}_{\textrm{C-N}}(r)$ partial PDFs, all 8000-atom models show two peaks at 1.25 and 1.43 {\AA} attributed to double C=N or triple C$\equiv$N bonds and to C-N single bonds, respectively. 
The SiCNH8000-Random model contains the highest amount of short C-N bonds, again due to its random initialization, whereas this bond type is prohibited in the setup of the ceramicNetworkBuilder program used to generate SiCNH8000-CNB, SiCNH8000-CL and SiCNH8000-CNB\_2200 initial configurations.
Nevertheless, few C-N bonds occurs in these models.\\

Similarly to N-N correlations, the main C-N peak around 3.05 {\AA} characterizes the C-(Si)-N distances of the SiC$_x$N$_{4-x}$ tetrahedral network, and the shoulder at lower $r$ values (around 2.6 {\AA}) corresponds to -Si-C-Si-N four-atom cycles. In this case, the narrow and symmetrical shape of the main peak in MLIP-MD models is a strong indication that artefactual environments present after model initialization continue to be eliminated as high-temperature MLIP-MD simulations extend far beyond typical time scales accessible to FPMD. \\

The Si-Si partial PDFs are similar for all four MLIP models. The only main difference can be seen in the SiCNH8000-CNB\_2200 model where the highest intensity peak at 3.2 {\AA}, encompassing the Si-(N)-Si and Si-(C)-Si correlations, is more intense and slightly narrower, explaining the achieved best match with the experimental PDF. In addition, SiCNH8000-CL and SiCNH8000-CNB\_2200 models show a slightly reduced intensity of the first peak at 2.43 {\AA} corresponding to Si-Si bonds. These results establish that the model that underwent the longest thermal cycle leads to an improved connectivity of the SiCN network as well as a reduction in the number of Si-Si bonds. Experimentally, Si-Si bonds have been already suggested by Choong Kwet Yive et al. \cite{choong1992thermogravimetric} although they are known to rearrange at high temperature to more stable Si-C-Si bridges via methylene insertion.\\

Finally, the Si-N and Si-C atomic pairs exhibit quite similar distributions with a very intense peak centered at 1.81 {\AA} and 1.93 {\AA}, corresponding to Si-N and Si-C bonds, respectively.\\

\begin{figure}[!ht]
\centering
\includegraphics[width=0.45\linewidth]{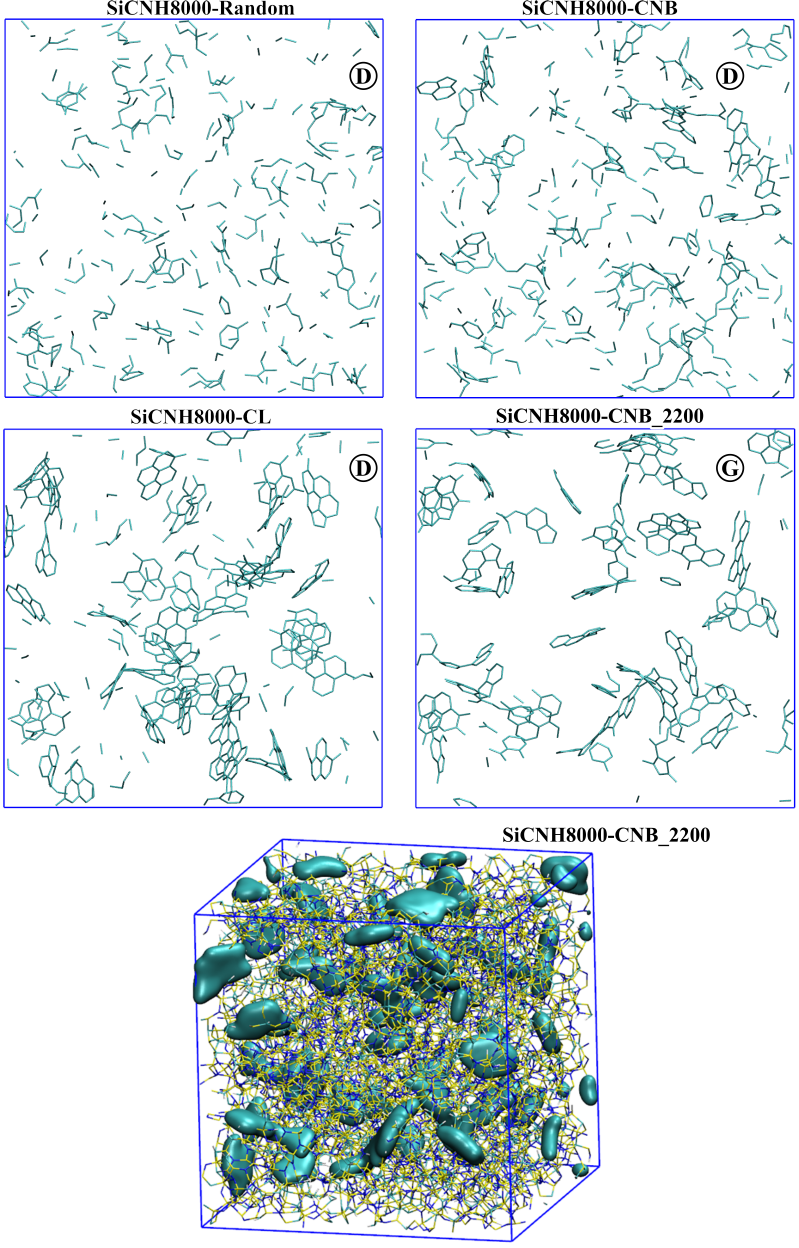}
\caption{(Top) Configurations of C-C bonds in the final snapshot of models SiCNH8000-Random, SiCNH8000-CNB, SiCNH8000-CL and SiCNH8000-CNB\_2200. Capital letters refer to the position on the thermal cycles shown in Figure \ref{fig:total_pdf}a. (Bottom) Perspective view of the final SiCNH8000-CNB\_2200 model, with carbon domains highlighted in QuickSurf iso-surface representation in VMD \cite{humphrey1996vmd}.}
\label{fig:snapshots_8000}
\end{figure}

Taken together, the above analysis show that the best two models, SiCNH8000-CL and in particular SiCNH8000-CNB\_2200, are consistent with the plausible presence of two separate phases: a carbon phase, containing sp$^2$-like C atoms, and a SiCN network. This can be confirmed by inspecting the C atoms organisation in the models. Figure \ref{fig:snapshots_8000} shows the projections of the simulation cells with only carbon atoms obtained at the end of the thermal annealing cycles. One can see that the best models, SiCNH8000-CL and SiCNH8000-CNB\_2200, show unambiguously the presence of more or less developed carbon sheets scattered in the cell, which is less the case of the two other MLIP models. Focusing on the SiCNH8000-CNB\_2200 model, the phase-separated carbon domain are actually embedded in the SiCN network as shown in the perspective view of Figure \ref{fig:snapshots_8000}. These observations are compatible with the formation of the well known free carbon phase in PDC materials and demonstrate the ability of the present MLIP-based methodology, associated with an extended size and time scales, to achieve realistic structural models with phase separations. 

\subsubsection{Local atomic environments}

Table \ref{tab:silicon_envs} shows the dominant Si local environments in the four 8000-atom models and Tables S4, S5, S6 and S7 in the supplementary materials, provide the complete list of local environments for all the  models. We find that, in the four models, between 87\% and 90\% of silicon atoms are 4-fold coordinated, as expected from a Si-based tetrahedral network.

\begin{table}[!ht]
\caption{Local silicon environments in the 8000-atom models, showing the fraction of main 4-fold coordination motifs. The full tables are provided in the supplementary materials paragraph S8.}
\label{tab:silicon_envs}
\centering
\begin{tabular}{lcccccc}
\toprule
Model      & \multicolumn{5}{c}{4-fold Si environments (\%)} \\
\cmidrule(lr){2-6}
 & Si-\ch{N4} & Si-\ch{CN3} & Si-\ch{C2N2} & Si-\ch{C3N} & Si-\ch{C4}  \\
\midrule
SiCNH8000-Random    & 5.2 & 19.0 & 24.7 & 13.9 & 3.2 \\
SiCNH8000-CNB       & 7.6 & 21.2 & 27.1 & 14.3 & 3.0  \\
SiCNH8000-CL        & 9.5 & 23.3 & 20.5 & 11.4 & 3.1  \\
SiCNH8000-CNB\_2200 & 8.3 & 23.3 & 27.9 & 14.0 & 2.3  \\
\bottomrule
\end{tabular}
\end{table}

Interestingly, silicon atoms are in majority bonded to nitrogen or carbon atoms forming mixed SiC$_x$N$_{4-x}$ tetrahedra. Unmixed Si-C$_4$ and Si-N$_{4}$ represent only around 3\% and less than 10\% for all models, respectively. Overall, there is no significant difference between the distribution of the silicon environments in the four models. One can only note that in SiCNH8000-CL the Si-C$_2$N$_2$ environments are less frequent, certainly due to the introduction of carbon sheets in the initial configuration which reduces \textit{de facto} the available carbon atoms that can participate in the formation of the SiCN network. 
Beside this, we find other less common silicon environments that involve hydrogen or other silicon atoms (see complete tables in the supplementary materials paragraph S8). Specifically, we note the existence of 3\% to 4\% environments such as Si-SiC$_2$N and Si-SiCN$_2$ in most of the models.\\

The analysis of the nitrogen local environments shows that it is predominantly 3-fold coordinated and essentially bonded to silicon atoms (in N-Si$_3$ or N-Si$_2$H environments) as expected in a SiCN network

\begin{table}[!ht]
\caption{Fractions of the main 3- and 4-fold carbon environments in the 8000-atom models. The full tables are provided in the supplementary materials paragraph S8.}
\label{tab:carbon_envs}
\centering
\begin{tabular}{lccccccccc}
\toprule
Model & \multicolumn{4}{c}{3-fold C environments (\%)} & \multicolumn{3}{c}{4-fold C environments (\%)} \\
\cmidrule(lr){2-5} \cmidrule(lr){6-8}
& C-\ch{Si3} & C-\ch{Si2C} & C-\ch{SiC2} & C-\ch{C3}  & C-\ch{Si4} & C-\ch{Si3H} & C-\ch{Si2H2}  \\
\midrule
SiCNH8000-Random & 4.7 & 7.6  & 6.3  & 3.5 & 21.1 & 22.9 & 6.4  \\
SiCNH8000-CNB    & 3.4 & 10.4 & 12.5 & 6.2 & 17.9 & 21.1 & 6.2   \\
SiCNH8000-CL     & 4.0 & 4.1 & 16.2 & 16.1 & 18.0 & 21.1 & 5.9   \\
SiCNH8000-CNB\_2200 & 4.9 & 1.0 & 12.3 & 15.0 & 24.2 & 20.8 & 7.5 \\
\bottomrule
\end{tabular}
\end{table}

Table \ref{tab:carbon_envs} shows the dominant 3- and 4-fold carbon local environments (see the full tables in the supplementary materials paragraph S8). We find that carbon atoms are primarily 4-fold ($\approx$ 52-58\%) and 3-fold ($\approx$ 37-47\%) coordinated. 
The 4-fold coordinated carbon atoms belong to the SiCN network whereas the 3-fold ones essentially belong to the free carbon domains. 
Focusing on the two best models, we observe that the the C-C$_3$ environment covers more than 15\% of the carbon atoms, reflecting the occurrence of carbon layers in contrast to the SiCNH8000-Random and SiCNH8000-CNB models. 
The carbon atoms located on the edges of these sheets maintain an sp$^2$ hybridization and are predominantly found in C-SiC$_2$ environments, building the connection between the matrix and carbon domains. These domains thus appear to be bound to the matrix, rather than solely interact via van der Waals interactions.
As for the C-Si$_2$C environments, while their fraction was around 10\%  SiCNH8000-CNB model, they almost disappear after the extended annealing of SiCNH8000-CNB\_2200 model, suggesting rather limited stability of this local configuration inside the Si-C-N-H system.

As previously mentioned, 4-fold carbon atoms located in the matrix are primarily bonded to Si atoms and show a significant tendency to incorporate hydrogen in their neighborhood with the nearly 20\% of C-Si$_3$H environments. Hydrogen atoms appears to be mainly bonded to carbon atoms of the matrix, playing a crucial role in stabilizing carbons that cannot form four bonds with silicon due to the constraints of the amorphous network. At the same time, the hydrogen atoms act as terminal atoms preventing the formation of a fully continuous random SiCN network. This, it turn, can possibly provide a higher flexibility to the amorphous network. It is worth-noting that only a relatively small fraction of hydrogen atoms are bonded to silicon in agreement with the lower stability of Si-H bonds in comparison to C-H ones \cite{seitz1996structural}.

\subsection{Free carbon domains characterization}
The local environments found in all models are globally similar with differences arising from the fraction of each environment. 
These models, due to their distinct initial configurations, may reflect different stages of material organization during the experimental pyrolysis. Indeed, the SiCNH8000-CNB model represents an intermediate stage, where the amorphous network develops few poorly organized carbon-rich domains. SiCNH8000-CNB\_2200 and SiCNH8000-CL models describe a more advanced stage, characterized by the formation and stabilization of larger carbon domains within the SiCN network, where carbon atoms forming C-C bonds generally adopt sp$^2$ hybridization, similar to the free carbon phase observed experimentally \cite{Chen2014Quantitative}. Finally, the SiCNH8000-Random model, while not representative of the early pyrolysis stages due to its non-physical random initialization, provides insights into the nucleation of the carbon phase and particularly the construction of aromatic rings from isolated atoms. In what follows we present the evolution of carbon domains by exploiting the four 8000-atom models.

\subsubsection{Evolution of C motifs}

\begin{figure}[!ht]
\centering
\includegraphics[width=0.78\linewidth]{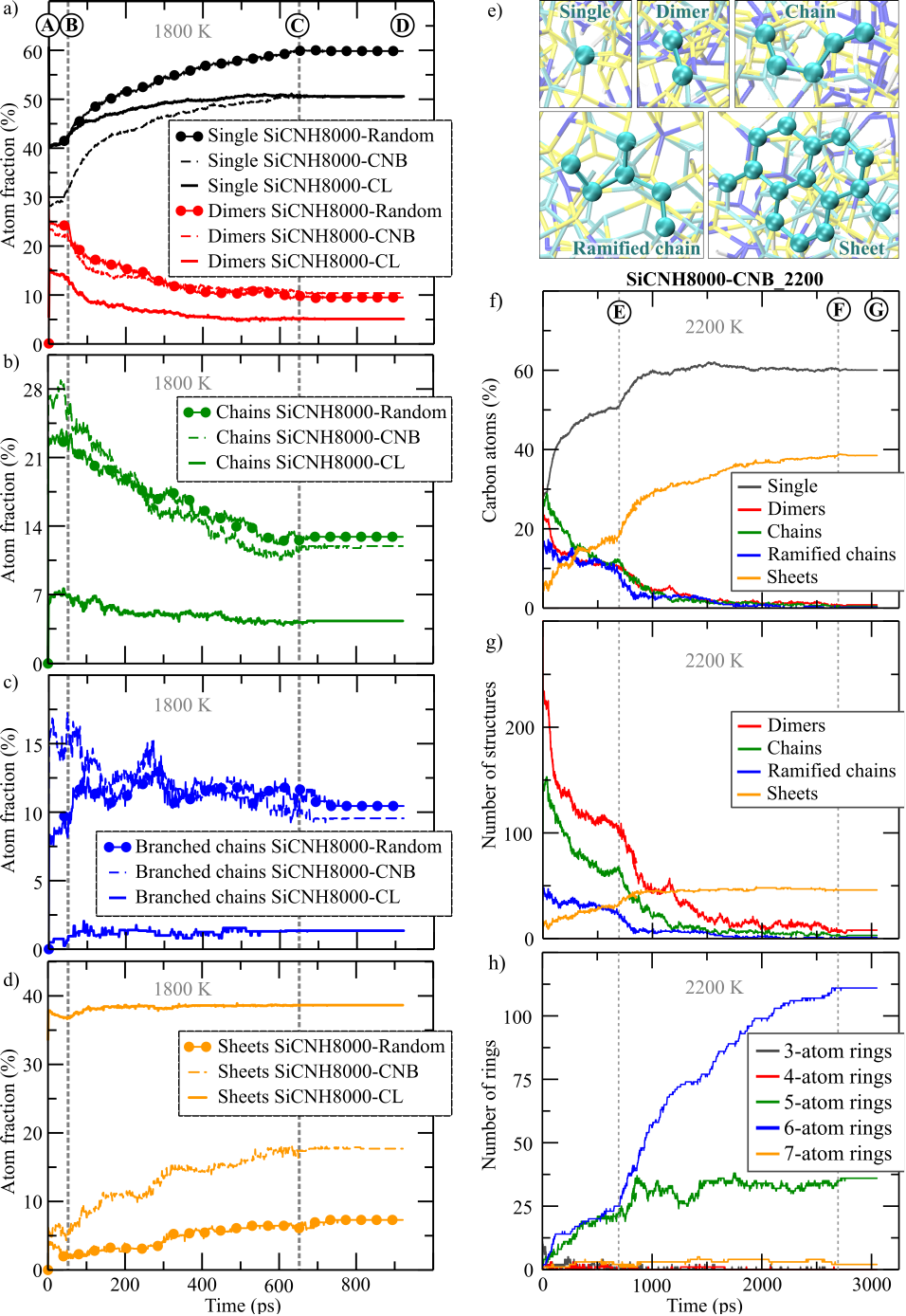}
\caption{Evolution of the proportion of carbon atoms in different environments (single, dimers, linear chains, branched chains and sheets) during the thermal cycle for the SiCNH8000-Random, SiCNH8000-CNB and SiCNH8000-CL (a-d, left panels) and SiCNH8000-CNB\_2200 (f). Evolution of (g) the number of specific carbon structures and of (h) the number of carbon rings with different sizes during the thermal cycle applied to SiCNH8000-CNB\_2200. The vertical dashed lines delimit the highest temperature plateau for each thermal cycles. Circled A-G capital letters indicate the timing configurations shown in Figure \ref{fig:total_pdf}a. The different types of carbon structures are illustrated (e) with examples extracted from the models.}
\label{fig:carbon_evolution}
\end{figure}

Figure S7 in the supplementary materials shows snapshots of the carbon domains of all models at different stages of the thermals cycles.
It reveals that, during the thermal annealing, carbon atoms undergo substantial organisation leading to the formation of dimers, chains and sheets. Specifically, we define five categories of carbon atoms (illustrated graphically in Fig. \ref{fig:carbon_evolution}e) characterizing typical structural carbon motifs and use them to track the system evolution during thermal annealing cycle.
The first category corresponds to "single" carbon atoms only bonded to silicon, nitrogen, and/or hydrogen, i.e. completely embedded in the SiCN network. The second category, labelled "dimers", corresponds to C-C pairs. The "chains" category describes linear chains of at least three carbon atoms, and "branched chains" category contains chains with at least one ramification. The final category, called "sheets" designates carbon structures containing at least one 3- to 7-atom carbon ring. 

Figure \ref{fig:carbon_evolution}a-d and \ref{fig:carbon_evolution}f shows that the number of single C atoms increases significantly in all four models over time, and becomes in all cases the main category at the end of the dynamics, with a fraction of approximately 50\% to 60\% in line with the formation of the SiCN network. Conversely, the fractions of atoms belonging to dimers and linear chains decrease steadily, with the most significant decrease occurring during the high-temperature plateaus, as observed visually in Figure S7 in the supplementary materials. The evolution of the fractions of atoms belonging to branched chains, on the other hand, depends on the specific model. In the SiCNH8000-Random and SiCNH8000-CL models, the fractions grow steadily during the first 105 ps of the annealing cycle before stabilizing around 11 and 3\%, respectively, whereas it fluctuates with a global decrease down to 9\% during the cooling phase in the SiCNH8000-CNB model. Finally, the fraction of carbon atoms belonging to sheets, already high by construction at the initial stage in model SiCNH8000-CL, is almost constant around 39\%, seemingly due to the lack of available carbon in the direct surroundings of the pre-constructed sheets \cite{boero2022atomistic}. In contrast, this fraction increases steadily during the high-temperature plateau, from 2 to 6\% in model SiCNH8000-Random and from 6 to around 17\% in model SiCNH8000-CNB. As for SiCNH8000-CNB\_2200 (Fig. \ref{fig:carbon_evolution}f,g) we observe a faster disappearance of dimers and linear/branched chains in favor of isolated atoms and sheets at the beginning of the 2200 K plateau. After t = 300 ps at this temperature, the growth of sheets and the isolated carbon atoms in the SiCN network exhibits a tendency to slow down.\\

We now focus on the evolution of carbon ring population in the models as shown in Figures S8 and \ref{fig:carbon_evolution}h. 
For SiNCH8000-Random and SiNCH8000-CNB models, we observe an increase of the number of 5 and 6-member rings over time, with the latter showing a faster growth rate. In addition, the 3 and 4-member rings almost disappear during the very first stage of annealing ($\approx$ 50 ps). In these systems, 7-member rings remain marginal during the annealing. The evolution in model SiNCH8000-CL differs significantly, due to the presence of 6-member carbon sheets in the initial configuration. During the dynamics, only 4 additional six-atom rings are formed at the periphery of pre-existing sheets. It is also notable that during the dynamics 9 five-atom rings form after 100 ps. These rings are found at the border of the already existent sheets, inducing a slight growth of these domains.\\

Coming to the best SiNCH8000-CNB\_2200 model, except at the beginning of the T = 2200 K plateau, we observe that the number of 5-member rings is constant (Fig. \ref{fig:carbon_evolution}h), whereas the number of 6-member rings undergoes a very rapid increase leading to the formation of carbon sheets dominated by 6-member rings. We note that, at the same time, dimers, chains and branched chains disappear, which suggests that they contribute to the growth of the extended carbon sheets. Interestingly, as the number of sheets remains constant, this trend implies a growth of the layers instead of the increase of their number (Fig. \ref{fig:carbon_evolution}g-h). At the end of the thermal annealing cycle, sheets representative of the free carbon phase gathers about 38\% of the carbon atoms (noticeably close to the amount of sp$^2$ carbon atoms in the model according to Table S7), while only 2\% persist inside dimers or chain-liked structures. The remaining 60\% of carbon atoms are involved in the formation of the SiCN network.
The PDC global composition can thus be decomposed into 40\% of free carbon and 60\% of C within the SiCN matrix with a composition \ch{Si32C15N24} (without accounting for H). This matrix composition is close to \ch{Si32C14N24}, consistent with a mixture of 70\% SiC and 30\% \ch{Si3N4} that can be expressed with the following equation:

\begin{equation}
\ch{Si32C25N24} \approx 10\ \ch{C} + 20\  \ch{(SiC)}_{0.7}(\ch{Si3N4})_{0.3}.
\end{equation}

This resulting chemically consistent phase separation further supports the ability of extended machine learning molecular dynamics at sufficiently high temperatures to properly model these complex SiCN systems.

\subsubsection{Nucleation and growth mechanisms}

\begin{figure}[!ht]
\centering
\includegraphics[width=0.7\linewidth]{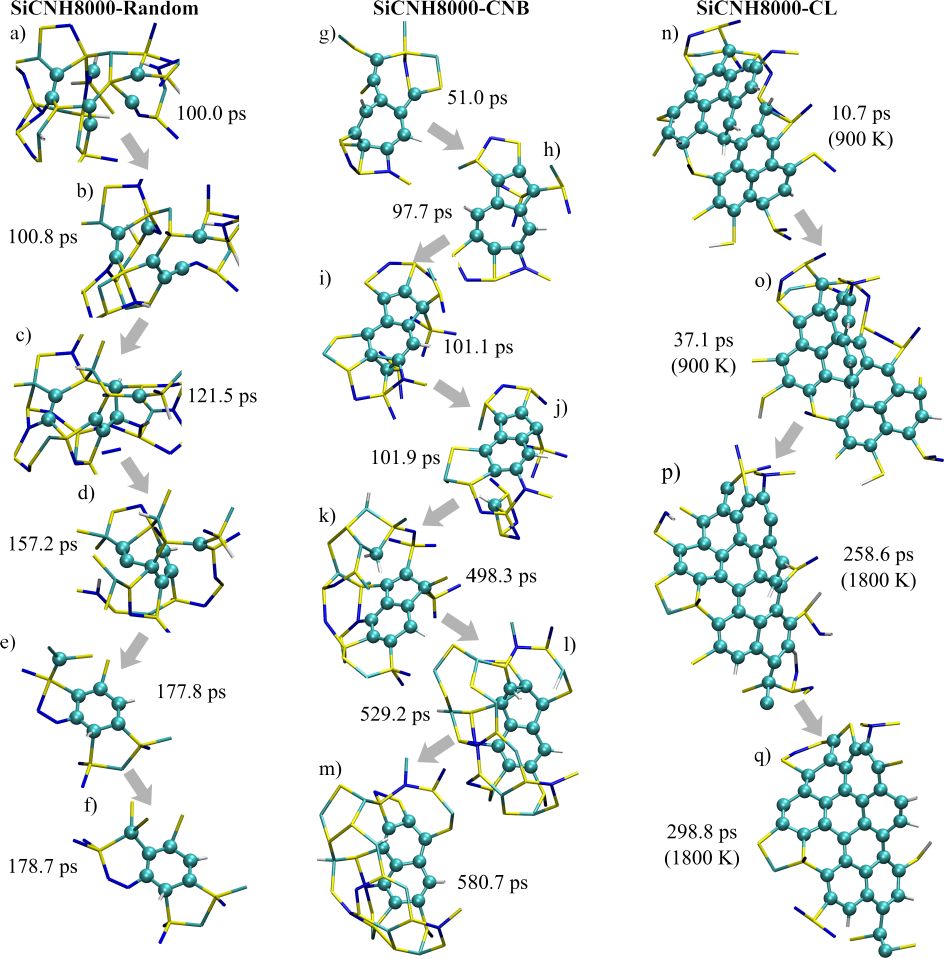}
\caption{(a-f) Atomic configurations illustrating the mechanism of aromatic ring formation from isolated carbon atoms in the SiCNH8000-Random model. The spheres represent carbon atoms that are part of the final ring, highlighting the progressive assembly of dispersed carbon atoms into organized aromatic structures. (g-m) Atomic configurations illustrating the mechanisms of 6-member aromatic ring formation from 5- and 7-member rings in the SiCNH8000-CNB model. The spheres represent carbon atoms that are part of the final ring, demonstrating the transformation pathways between different ring structures. (n-q) Atomic configurations illustrating the coalescence of two carbon sheets in the SiCNH8000-CL model. The spheres represent carbon atoms that are part of the final sheet, demonstrating the complex process of sheet merging and structural reorganization. Yellow, blue and white lines represent bonds made by the silicon, nitrogen and hydrogen atoms, respectively.}
\label{fig:ring-formation-mech}
\end{figure}

Figure \ref{fig:ring-formation-mech}a-f show local atomic configurations in the SiCNH8000-Random model at different times during the T = 1800 K plateau, highlighting the formation of an aromatic C ring. Starting from single carbon atoms and carbon dimers initially dispersed in the amorphous ceramic network (Figure \ref{fig:ring-formation-mech}a), we observe the progressive formation of a linear chain of 6 atoms through the coalescence of the initial structures (Figure \ref{fig:ring-formation-mech}b-d). The formation of new C-C bonds allowing the growth of a carbon chain and eventually a ring is enabled by the breaking of Si-C or C-N bonds, with Si atoms bonding to nearby C or N atoms and N atoms to nearby Si atoms. The 6-atom linear chain (Figure \ref{fig:ring-formation-mech}d) closes to form a 6-member ring (Figure \ref{fig:ring-formation-mech}e), which later captures an additional carbon atom at its periphery (Figure \ref{fig:ring-formation-mech}f), potentially initiating further sheet growth through the construction of an additional adjacent ring. Similar trends were reported by Imoto et al. using metadynamics simulations, where carbon atoms first form polymer-like chains, followed by rings, and eventually hexagonal, graphene-like structures~\cite{boero2022atomistic}.\\

As highlighted in Figures S8 and \ref{fig:carbon_evolution}, rings containing 3, 4, 5 or 7 carbon atoms occasionally appear during the dynamics, among which only 5-member rings may be preserved throughout the annealing cycle. These imperfect cycles play an important transient role in the formation of six-atom aromatic cycles and the growth of carbon sheets, as illustrated in Figure \ref{fig:ring-formation-mech}g-m, showing atomic configurations extracted from the thermal cycle applied to the SiCNH8000-CNB model. Configuration \ref{fig:ring-formation-mech}g, extracted at the beginning of the T = 1800 K plateau (t = 51 ps), contains a 7-atom carbon ring with two lateral branches made of 1 and 2 C atoms, which eventually bind to form an adjacent 5-atom ring (Figure \ref{fig:ring-formation-mech}h). Configurations \ref{fig:ring-formation-mech}i,j show one possible mechanism by which an aromatic 6-member ring is formed from a 7-atom ring. One of the C atoms is ejected from the initial ring via the formation of an evanescent 3-member ring wherein the (later) ejected C atom first forms a C-C$_2$Si$_2$ tetrahedron. It then quickly leaves the vicinity of the carbon sheet by bonding to a third Si atom and one hydrogen atom to integrate the SiCN network, like the majority of carbon atoms in the SiCNH8000-CNB model (see Table \ref{tab:carbon_envs}).

The opposite mechanism is observed around the 5-atom ring later during the thermal cycle. Specifically, a C atom sitting in a C-CSiH$_2$ tetrahedron, eventually positioned above two atoms of the 5-atom ring (Figure \ref{fig:ring-formation-mech}k), forms a 3-C ring (Figure \ref{fig:ring-formation-mech}l) similar to the one observed in Figure \ref{fig:ring-formation-mech}i, which then intercalates into the carbon ring to convert it into a stable aromatic 6-member cycle (Figure \ref{fig:ring-formation-mech}m).\\

While carbon sheets growth mainly occurs through the capture of single atoms or dimers, the coalescence of nearby sheets has also been observed in the SiCNH8000-CL model, as illustrated in Figure \ref{fig:ring-formation-mech}n-q. Configuration \ref{fig:ring-formation-mech}n, obtained after 10.7 ps (at T = 900 K), shows two carbon sheets made of 4 and 3 rings, and connected by a  shared lateral C-C bond. A 1-C-atom branch located at the periphery of the 3-ring sheet binds, in configuration \ref{fig:ring-formation-mech}o, with the 4-ring sheet to form a new 6-atom cycle, wherein three atoms are shared with a 7-atom ring created by the approach of the two initial sheets. This new extended carbon structure contains C atoms located above its main plane, which then progressively move to its periphery to obtain a nearly-planar structure (Figure \ref{fig:ring-formation-mech}p). A new aromatic cycle is finally formed, leading to a single large sheet composed of 7 6-member and 1 7-member rings (Figure \ref{fig:ring-formation-mech}q).\\

Taken together Figures \ref{fig:ring-formation-mech}a-f, \ref{fig:ring-formation-mech}g-m, and \ref{fig:ring-formation-mech}n-q highlight the mechanisms of nucleation and growth of free carbon domains in Si$_{32}$C$_{25}$N$_{24}$H$_{19}$ models, wherein ring formation appears to be a strongly defect-mediated process, with short-lived 3, 4, 5 and 7-member rings playing key roles. 
This growth mechanism is inline with recent metadynamics simulation of carbon layer formation on SiC(0001) surface, where it was shown that under-coordinated C reorganize in lines, rings and eventually aromatic seeds to ultimately form graphene \cite{boero2022atomistic}.\\

The extension of the dynamics and the temperature rise applied to the SiCNH8000-CNB\_2200 model allowed to continue the growth and the structuring of free carbon domains initiated in the SiCNH8000-CNB model. During this high temperature dynamics, nearly all dimers and carbon chains disappear, leaving carbon atoms in a phase separated configuration: isolated in the SiCN network or belonging to graphitic sheets. 

\section{Conclusion}

In this work, we present a generic methodology for constructing a highly accurate machine learning interatomic potential for polymer-derived silicon carbonitride (Si-C-N-H) ceramics. To capture the full complexity of amorphous Si-C-N-H systems, our approach combines a meticulously constructed training database containing FPMD amorphous models and diversified with crystal structure prediction models, together with advanced machine learning architectures. The obtained MACE potential shows an accuracy of 12.4 meV/atom for energies and 149 meV/{\AA} for forces, with strong transferability across diverse structural environments and compositions.\\

Large-scale molecular dynamics simulations of 8000-atom systems of composition Si$_{32}$C$_{25}$N$_{24}$H$_{19}$ with various initial configurations and thermal cycles led to four PDC models. Compared to PDF experiments,
the best model is unambiguously the one that spontaneously evolves towards a clear phase separation, characterized by dispersed carbon sheets embedded in an SiCN amorphous phase, starting from an appropriately-tuned initialization and given a long enough simulation time at a temperature deliberately higher than experimental pyrolisis temperature.
For the considered PDC composition, we find that about 40\% of the carbon atoms belong the sheets, while the remaining 60\% carbon fraction participate to the formation of the SiCN network. This latter is mainly built of 4-fold silicon atoms bonded to both nitrogen and carbon atoms, 4-fold carbon atoms mainly bonded to silicon atoms and finally 3-fold nitrogen atoms. The occurrence of this biphasic system, requires a long thermal annealing at high temperatures (T = 2200 K). It brings the calculated atomic pair distribution function into exceptional agreement with experimental data, particularly in the medium-range order, suggesting a similar carbon distribution in the real sample.\\ 

The structuring of the free carbon phase goes along with the disappearance of carbon dimers and small carbon chains as well as the growth of carbon sheets containing mainly aromatic rings.
The nucleation and growth of these carbon sheets are mediated by defective carbon rings, which serve as intermediate structures that facilitate the formation of stable 6-member aromatic cycles through the capture of additional carbon atoms or the ejection of excess ones to optimize local bonding configurations. \\

The convergence of our models toward experimentally observed structures, regardless of their initial configurations, underscores the thermodynamic robustness of our potential and its ability to capture the intrinsic structural features of these complex ceramics and the mechanisms of free carbon phase growth. Beyond the specific composition studied in this work, this methodology can be applied to carbon rich PDC systems to study the formation of extended turbostratic carbon structures. \\

\noindent\textbf{\Large Acknowledgements}\\

Calculations were performed by using resources from Grand Equipement National de Calcul Intensif (GENCI, grants no. AX0913426 and AX0910832). Computational resources provided by the computing facilities Mésocentre de Calcul Intensif Aquitain (MCIA) of the Université de Bordeaux and of the Université de Pau et des Pays de l’Adour. We acknowledge funding Agence national de recherche under the RECIFE ANR-DFG project (Grant Number ANR-21-CE08-0036-01).\\

\noindent\textbf{\Large Data Availability}\\

The Si-C-N-H database, converged MLIP and sample 8000-atom models will be made available at ....\\

\bibliographystyle{apsrev4-2}
\bibliography{short-bib}

\pagebreak
\clearpage

\setcounter{equation}{0}
\setcounter{figure}{0}
\setcounter{table}{0}
\setcounter{page}{1}
\setcounter{section}{0}
\makeatletter
\renewcommand{\theequation}{S\arabic{equation}}
\renewcommand{\thefigure}{S\arabic{figure}}
\renewcommand{\thetable}{S\arabic{table}}
\renewcommand{\thesection}{S\arabic{section}}

\begin{center}
\textbf{\large Supplementary Materials: Modeling phase separation in polymer-derived silicon carbonitride ceramics through extended machine learning molecular dynamics}
\vspace{0.25cm}\\
{\large Fabien Mortier, Sylvian Cadars, Olivier Masson, Mauro Boero, Guido Ori, Yun Wang, Samuel Bernard, and Assil Bouzid.}
\end{center}

\section{Additional periodic models}
Three additional periodic models, of composition \ch{Si32C25N24H19}, containing 400 atoms are constructed to enrich the database of amorphous \ch{SiCNH} systems. 

The first model is produced using Born-Oppenheimer Molecular Dynamics (BOMD) as implemented in the CP2K suite of programs \cite{Kuhne2020CP2K}. The calculations were conducted in the NVT ensemble in which the temperature is monitored by a Nosé-Hoover thermostat chain \cite{Nose1984Canonical, Nose1984Unified, Hoover1985Canonical, martyna1992nose}. A timestep of 1 fs was used to integrate the equations of motion. 
The initial atomic configuration is generated with the ceramicNetworkBuilder program, employing bonding and coordination probabilities identical to those used for the SiCNH200 and SiCNH400-C models in Ref. \cite{mortierfpmd2025}. Specifically, Si atoms have a 0\% probability of bonding with Si, a 10\% probability of bonding with C, and a 90\% probability of bonding with N. The coordination probabilities are set such that C atoms have a 75\% probability of 3-coordination, while Si atoms have a 100\% probability of 4-coordination and N atoms have a 100\% probability of 3-coordination. The thermal cycle applied during BOMD simulations includes the following stages: 4 ps at 600 K, 7 ps at 1300 K, 7 ps at 1800 K, 10 ps at 1300 K, 5 ps at 900 K, 5 ps at 600 K, and 15 ps at 300 K. This cycle allows the system to explore a wide temperature range, ensuring thorough relaxation.

The second model follows the same procedure as the SiCNH400-CL-3 model in Ref. \cite{mortierfpmd2025}. Initially, three graphene-like sheets, each consisting of 4 carbon rings, are inserted into a periodic cell. The ceramicNetworkBuilder program code is then used to construct the amorphous \ch{SiCNH} network around these sheets, achieving a final composition of \ch{Si32C25N24H19}. The bonding probabilities for this model are set such that Si atoms have a 0\% probability of bonding with Si, a 25\% probability of bonding with C, and a 75\% probability of bonding with N. C and N atoms have a 100\% probability of 3-coordination, and Si atoms have a 100\% probability of 4-coordination. After generating the initial configurations, a geometry optimization step is performed, followed by a thermal cycle using the second-generation Car-Parrinello molecular dynamics (SGCPMD) \cite{Kuhne2007SGCPMD, Kuhne2014SGCPMD2}. The thermal cycle for this model includes: 4 ps at 300 K, 4 ps at 600 K, 20 ps at 900 K, 10 ps at 1100 K, 12 ps at 900 K, 12 ps at 600 K, and 30 ps at 300 K. 

The third model, named SiCNH400-CL-4, is generated using a procedure similar to that of SiCNH400-CL-3 model in Ref. \cite{mortierfpmd2025}. The only difference is that it incorporates two sheets of seven carbon rings dispersed within the amorphous \ch{SiCNH} network. The final atomic configuration is achieved after applying the following second-generation Car-Parrinello molecular dynamics annealing thermal cycle: 4 ps at 300 K, 4 ps at 600 K, 20 ps at 900 K, 10 ps at 1100 K, 12 ps at 900 K, 12 ps at 600 K, and 30 ps at 300 K. 

\section{Thermal cycle used for generating the high-temperature configurations}

\begin{figure}[!ht]
\centering
\includegraphics[width=0.45\linewidth]{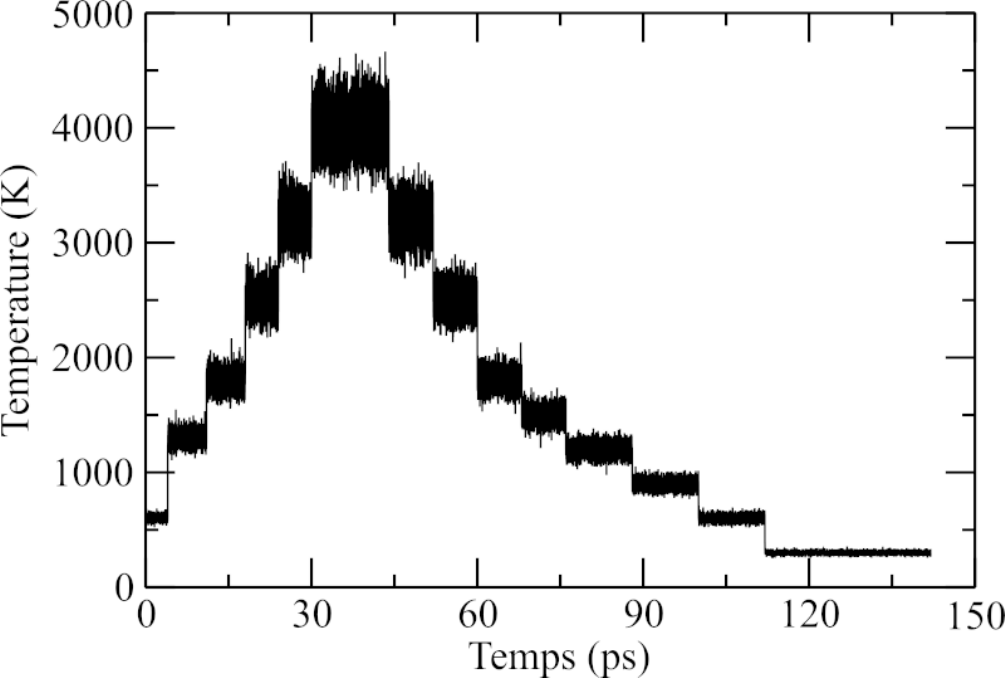}
\caption{Thermal cycle used for generating the high-temperature configurations in the SiCNH-periodic-4000K class.}
\label{fig:thermal_4000K}
\end{figure}

\section{Surface models generation protocol}
Starting from periodic models, the process of building surfaces for systems of varying sizes and molecular dynamics (MD) methods begins by cutting the periodicity of the simulation cell in the $z$ space direction, through the insertion of a layer of void of approximately 20 \AA{}. 
After a DFT-geometry optimization step, the surfaces are stabilized using a thermal cycle. For systems containing 200 atoms, BOMD is employed, while SGCPMD is used for models with 400 atoms. Both approaches share a consistent time step of 1 femtosecond for integrating the equations of motion.

In the BOMD framework, simulations are conducted within an NVT ensemble, where temperature control is achieved using a Nosé-Hoover thermostat. For SGCPMD, the conservation of temperature and total system energy relies on adjusting the Langevin friction coefficients, denoted as $\gamma_L$ and $\gamma_D$. The coefficient $\gamma_L$ is fixed at 0.1 fs\textsuperscript{-1}, while $\gamma_D$ is adjusted between 0.018 and 0.020 fs\textsuperscript{-1}, depending on the temperature target, to ensure the conservation of the total energy. 

The thermal cycle applied to relax the ceramic surfaces follows a gradual sequence: 5 ps at T = 300 K, 5 ps at T = 600 K, 5 ps at T = 900 K, followed by 10 ps at T = 1300 K, then descending in steps of 5 ps at T = 900 K and T = 600 K, and finally 20 ps at T = 300 K. This protocol allows for exploration across a wide temperature range, promoting optimal relaxation of the structures while avoiding abrupt transitions that could compromise their stability. 

\section{Crystal structure prediction}

Table S1 gives the main parameters used to perform multi-stage relaxation and energy calculation with the VASP code for every structure generated by the crystal structure prediction algorithm. The Perdew-Burke-Ernzerhof (PBE) exchange and correlation functional \cite{Perdew1996PBE} was used at all stages, with van der Waals interactions described with Grimme's DFT-D3 method with zero-damping function \cite{Grimme2010DFT-D3}. The maximum spacing between k-points in the table uses the VASP definition, wherein the number of k-points in a given direction $i$ of the reciprocal space is given by $N_i = \max(1, \lceil | \mathbf{b}_i | 2\pi / k_{spacing} \rceil)$, which involves a $2\pi$ factor with respect to the definition used in USPEX. This gives dense k-point meshes able to precisely describe potentially non-insulating systems in the last stages. Similar parameters were used for fixed-low-density CSP runs, with all relaxation steps performed with free shape but fixed volume, with a cutoff energy sequence of 350, 400 and 450 eV, (and the same value of 520 eV for the final energy calculation). 

\begin{table}[!ht]
\centering
\caption{Summary of computational parameters for atomic relaxation and energy calculations.}
\begin{tabular}{cccccccc}
\toprule
\makecell{Step} &
\makecell{Type} &
\makecell{Elect.-energy \\ threshold \\ (eV)} &
\makecell{Ionic-energy \\ threshold \\ (eV)} &
\makecell{Volume/\\shape} &
\makecell{Cutoff \\ energy \\ (eV)} &
\makecell{K-point \\ spacing \\ (\AA{}$^{-1}$)} &
\makecell{Gaussian \\ smearing \\ (eV)} \\
\midrule
1    & relaxation & $10^{-3}$ & $10^{-2}$ & fixed   & 300 & 0.63 & 0.08 \\
2    & relaxation & $10^{-4}$ & $10^{-3}$ & free    & 390 & 0.38 & 0.06 \\
3    & relaxation & $10^{-5}$ & $10^{-4}$ & free    & 520 & 0.19 & 0.04 \\
4    & energy     & $10^{-6}$ & N.A.     & N.A.    & 520 & 0.12 & 0.04 \\
\bottomrule
\end{tabular}
\end{table}

Different selection procedures were combined to select 100 to 200 structures in each USPEX run. Some structures were selected randomly, while others were selected by first truncating the energy range of all structures in a run into fragments and then trying to pick in each energy segment a structure as distinct as possible (in the sense of Valle-Oganov distances \cite{Valle2010Fingerprint}) from the one selected in the energy-segment below as illustrated in Figures S2 and S3.  

\begin{figure}[!ht]
\centering
\includegraphics[width=0.6\linewidth]{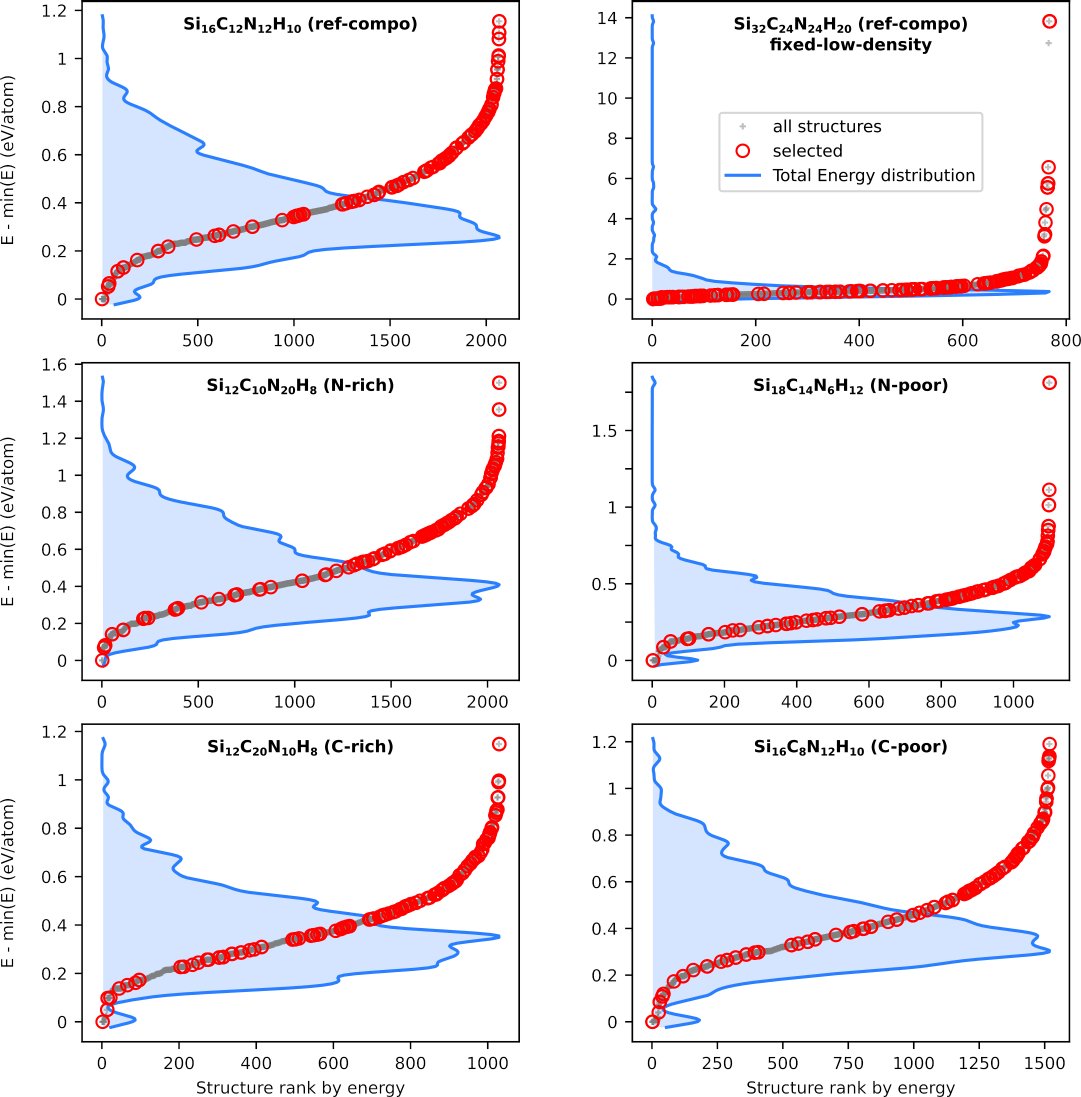}
\caption{Relative energies of structures explored within different CSP runs (grey crosses), with the corresponding energy distribution shown in blue. Structures selected to feed the MLIP training database are shown as red circles.}
\label{fig:csp_energies_1}
\end{figure}

\begin{figure}[!ht]
\centering
\includegraphics[width=0.6\linewidth]{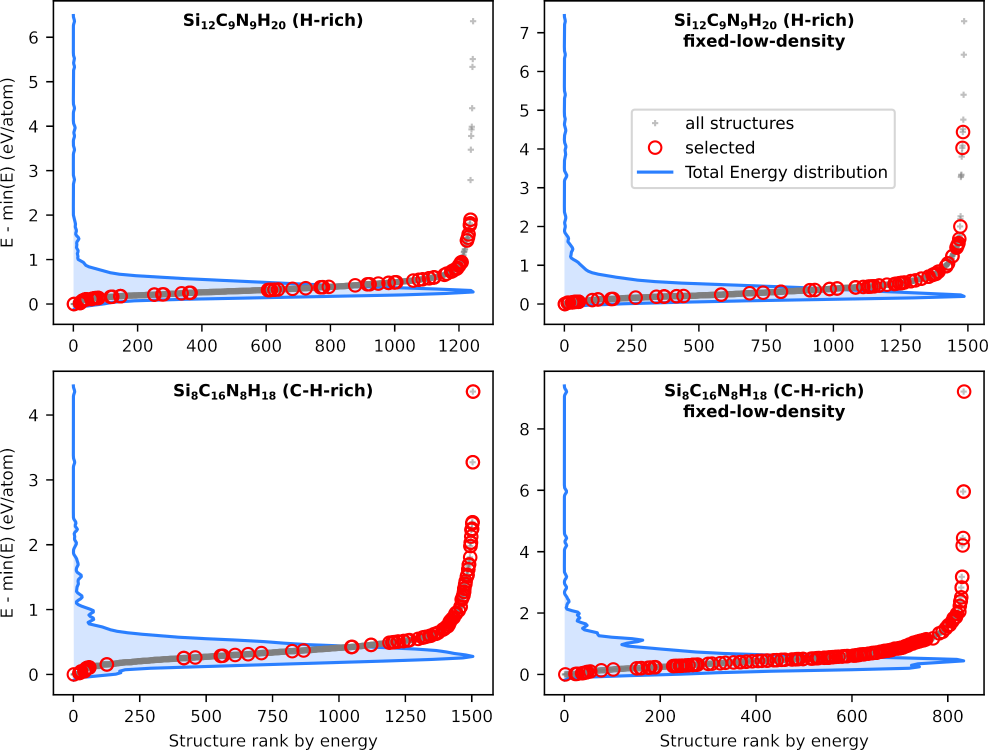}
\caption{Relative energies of structures explored within other CSP runs (grey crosses), with the corresponding energy distribution shown in blue. Structures selected to feed the MLIP training database are shown as red circles.}
\label{fig:csp_energies_2}
\end{figure}

\clearpage
\newpage
\section{Model training}

\begin{figure}[!ht]
\centering
\includegraphics[width=0.75\linewidth]{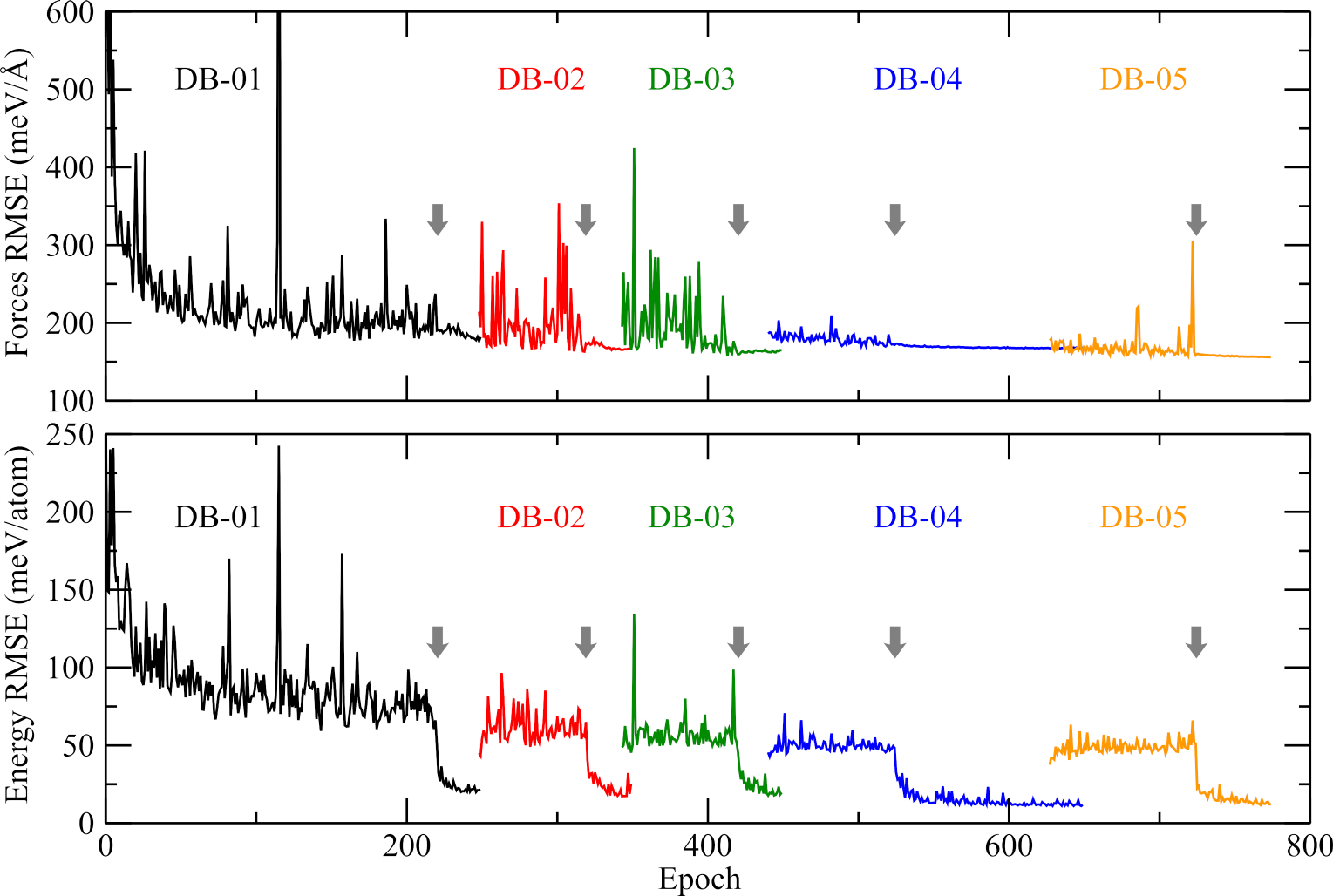}
\caption{Evolution of root mean square errors (RMSE) for energies and forces on the validation set during the training process. Arrows indicate switches from the force-optimization to the energy-optimization stage.}
\label{fig:training_RMSE_convergence}
\end{figure}

\section{Initial parameters of the 8000 atoms models}

Initial configurations of models SiCNH8000-CNB and SiCNH8000-CL were generated using the ceramicNetworkBuilder program \cite{mortierfpmd2025} with target bonding and coordination number probabilities given in Tables S\ref{CNB_Bond_Probas_SiCNH8000} and S\ref{CNB_Coord_Probas_SiCNH8000}.
These parameters allow the construction of an amorphous network composed of Si, C and N atoms. Hydrogen atoms are inserted afterwards following the target SiCNH composition to fill incomplete atomic local environments as in our previous work \cite{mortierfpmd2025}.

\begin{table}[!ht]
    \centering
    \caption{Target bonding probabilities used in CeramicNetworkBuilder to generate the initial configurations of SiCNH8000-CNB and SiCNH8000-CL models.}
    \begin{tabular}{|c|c|ccc|}
    \toprule
    Model & Neighbor type & \multicolumn{3}{c|}{Central atom type} \\
    {} & {} & Si & C & N \\
    \midrule
    SiCNH8000-CNB & 2 & 0 & 0 & 0 \\
    {} & 3 & 0 & 20 & 100 \\
    {} & 4 & 100 & 80 & 0 \\
    \midrule
    SiCNH8000-CL & 2 & 0 & 0 & 0 \\
    {} & 3 & 0 & 0 & 100 \\
    {} & 4 & 100 & 100 & 0 \\
    \bottomrule
    \end{tabular}
    \label{CNB_Bond_Probas_SiCNH8000}
\end{table}

\begin{table}[!ht]
    \centering
    \caption{Target coordination number probabilities used in CeramicNetworkBuilder to generate the initial configurations of SiCNH8000-CNB and SiCNH8000-CL models.}
    \begin{tabular}{|c|c|ccc|}
    \toprule
    Model & Neighbor type & \multicolumn{3}{c|}{Central atom type} \\
    {} & {} & Si & C & N \\
    \midrule
    SiCNH8000-CNB & Si & 0 & 80 & 100 \\
    {} & C & 40 & 20 & 0 \\
    {} & N & 60 & 0 & 0 \\
    \midrule
    SiCNH8000-CL & Si & 0 & 100 & 100 \\
    {} & C & 25 & 0 & 0 \\
    {} & N & 75 & 0 & 0 \\
    \bottomrule
    \end{tabular}
    \label{CNB_Coord_Probas_SiCNH8000}
\end{table}

\begin{figure}[!ht]
\centering
\includegraphics[width=0.99\linewidth]{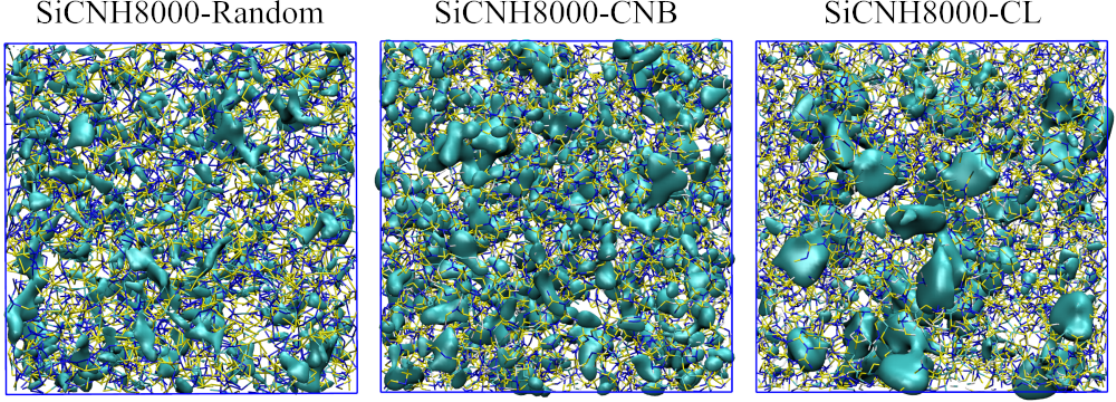}
\caption{Initial configurations of the SiCNH8000-Random, SiCNH8000-CNB and SiCNH8000-CL models. Turquoise regions indicate carbon-rich domains highlighted in QuickSurf iso-surface representation in VMD \cite{humphrey1996vmd}. Yellow, blue and white lines represent bonds made by the silicon, nitrogen and hydrogen atoms, respectively.}
\label{fig:initial_configs}
\end{figure}

\clearpage
\section{PDFs scaling and comparison}

\begin{figure}[!ht]
\centering
\includegraphics[width=0.46\linewidth]{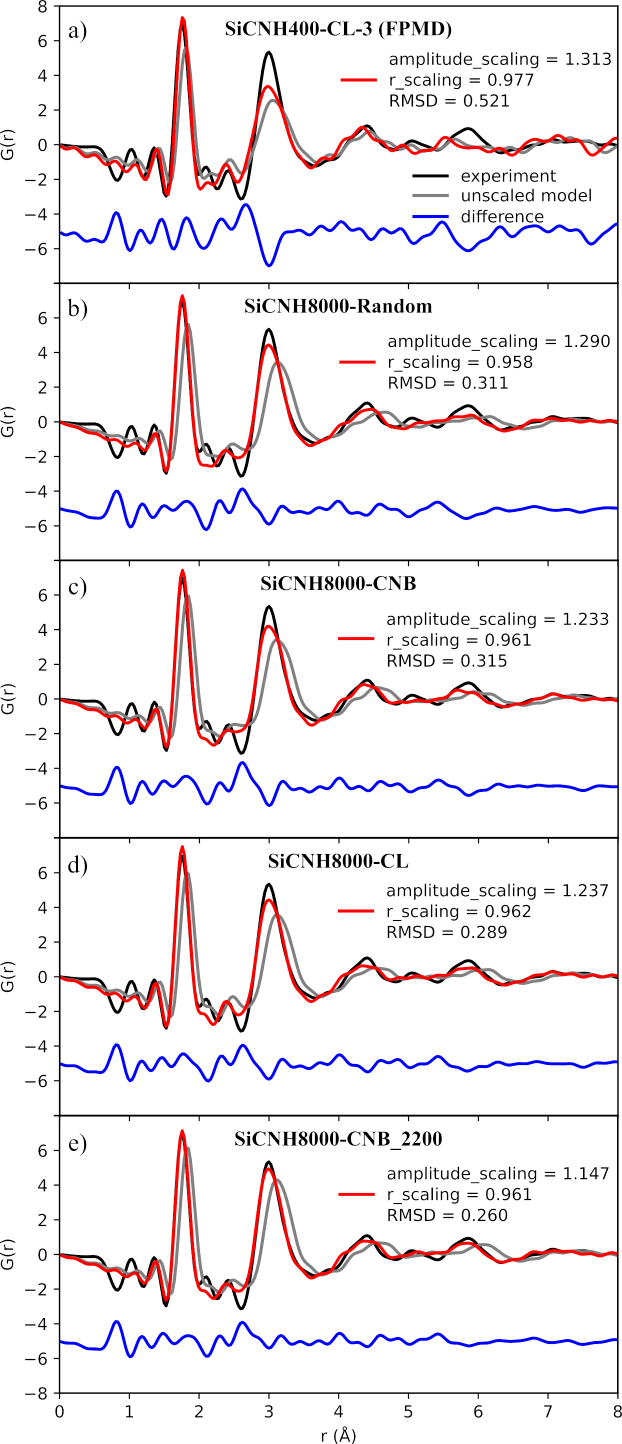}
\caption{Total X-ray scattering PDF data simulated for all SiCNH PDC models studied in this work, and compared to experimental data (in black). Data in grey are the raw PDF data. Data in red correspond to simulated PDF data wherein an amplitude scaling factor and a r-scaling factor have been optimized to best match to the experiment. The corresponding scaling factor and RMSD are indicated in the legend. Difference between scaled-optimized simulated data and experiment are shown below, in blue.}
\label{fig:PDF_scaling_fits}
\end{figure}

\clearpage
\newpage
\section{Atomic local environments in the MLIP 8000-atom models}

\begin{table}[!ht]
\small
\caption{Local atomic environments of the SiCNH8000-Random model. The "Other" sections compile environments whose individual fraction is below 3\%.}
\label{local_environments_SiCNH8000-Pack}
\begin{tabular*}{\columnwidth}{@{\extracolsep\fill}lcccccccccc@{\extracolsep\fill}}
\toprule
\multicolumn{10}{c}{Proportion [\%]}\\
\makecell{Nombre de\\coordinence} & \multicolumn{2}{c}{l=1} & \multicolumn{2}{c}{l=2} & \multicolumn{2}{c}{l=3} & \multicolumn{2}{c}{l=4} & \multicolumn{2}{c}{l$\geq$5}\\
\midrule
\makecell{Si} & \multicolumn{2}{c}{0.1} & \multicolumn{2}{c}{0.7} & \multicolumn{2}{c}{8.1} & \multicolumn{2}{c}{87.1} & \multicolumn{2}{c}{3.8}\\
\midrule
{} & Other & 0.1 & Other & 0.7 & Other & 8.1 & Si-\ch{C2N2} & 24.7 & Other & 3.8\\
{} & {} & {} & {} & {} & {} & {} & Si-\ch{CN3} & 19.0 & {} & {}\\
{} & {} & {} & {} & {} & {} & {} & Si-\ch{C3N} & 13.9 & {} & {}\\
{} & {} & {} & {} & {} & {} & {} & Si-\ch{SiC2N} & 5.9 & {} & {}\\
{} & {} & {} & {} & {} & {} & {} & Si-\ch{N4} & 5.2 & {} & {}\\
{} & {} & {} & {} & {} & {} & {} & Si-\ch{SiCN2} & 4.0 & {} & {}\\
{} & {} & {} & {} & {} & {} & {} & Si-\ch{C4} & 3.2 & {} & {}\\
{} & {} & {} & {} & {} & {} & {} & Other & 11.2 & {} & {}\\
\midrule
\makecell{C} & \multicolumn{2}{c}{0} & \multicolumn{2}{c}{3.2} & \multicolumn{2}{c}{37.4} & \multicolumn{2}{c}{56.9} & \multicolumn{2}{c}{0}\\
\midrule
{} & {} & {} & Other & 3.2 & C-\ch{Si2C} & 7.6 & C-\ch{Si3H} & 22.9 & {} & {}\\
{} & {} & {} & {} & {} & C-\ch{SiC2} & 6.3 & C-\ch{Si4} & 21.1 & {} & {}\\
{} & {} & {} & {} & {} & C-\ch{Si3} & 4.7 & C-\ch{Si2H2} & 6.4 & {} & {}\\
{} & {} & {} & {} & {} & C-\ch{C2N} & 4.1 & Other & 9.2 & {} & {}\\
{} & {} & {} & {} & {} & C-\ch{C3} & 3.5 & {} & {} & {} & {}\\
{} & {} & {} & {} & {} & C-\ch{SiCN} & {3.3} & {} & {} & {} & {}\\
{} & {} & {} & {} & {} & Other & 7.9 & {} & {} & {} & {}\\
\midrule
\makecell{N} & \multicolumn{2}{c}{3.7} & \multicolumn{2}{c}{7.8} & \multicolumn{2}{c}{86.5} & \multicolumn{2}{c}{2.0} & \multicolumn{2}{c}{0}\\
\midrule
{} & Other & 3.7 & N-\ch{SiC} & 5.1 & N-\ch{Si3} & 61.0 & Other & 2.0 & {} & {}\\
{} & {} & {} & Other & 2.7 & N-\ch{Si2H} & 14.9 & {} & {} & {} & {}\\
{} & {} & {} & {} & {} & N-\ch{Si2C} & 7.0 & {} & {} & {} & {}\\
{} & {} & {} & {} & {} & Other & 3.6 & {} & {} & {} & {}\\
\midrule
\makecell{H} & \multicolumn{2}{c}{99.9} & \multicolumn{2}{c}{0} & \multicolumn{2}{c}{0} & \multicolumn{2}{c}{0} & \multicolumn{2}{c}{0}\\
\midrule
{} & H-C & 66.2 & {} & {} & {} & {} & {} & {} & {} & {}\\
{} & H-N & 22.6 & {} & {} & {} & {} & {} & {} & {} & {}\\
{} & H-Si & 11.0 & {} & {} & {} & {} & {} & {} & {} & {}\\
{} & H-H & 0.1 & {} & {} & {} & {} & {} & {} & {} & {}\\
\bottomrule
\end{tabular*}
\end{table}

\clearpage
\newpage

\begin{table}[!ht]
\small
\caption{Local atomic environments of the SiCNH8000-CNB model. The "Other" sections compile environments whose individual fraction is below 3\%.}
\label{local_environments_SiCNH8000-CNBD}
\begin{tabular*}{\columnwidth}{@{\extracolsep\fill}lcccccccccc@{\extracolsep\fill}}
\toprule
\multicolumn{10}{c}{Proportion [\%]}\\
\makecell{Nombre de\\coordinence} & \multicolumn{2}{c}{l=1} & \multicolumn{2}{c}{l=2} & \multicolumn{2}{c}{l=3} & \multicolumn{2}{c}{l=4} & \multicolumn{2}{c}{l$\geq$5}\\
\midrule
\makecell{Si} & \multicolumn{2}{c}{0.1} & \multicolumn{2}{c}{0.6} & \multicolumn{2}{c}{5.7} & \multicolumn{2}{c}{88.8} & \multicolumn{2}{c}{4.7}\\
\midrule
{} & Other & 0.1 & Other & 0.6 & Other & 5.7 & Si-\ch{C2N2} & 27.1 & Other & 4.7\\
{} & {} & {} & {} & {} & {} & {} & Si-\ch{CN3} & 21.2 & {} & {}\\
{} & {} & {} & {} & {} & {} & {} & Si-\ch{C3N} & 14.3 & {} & {}\\
{} & {} & {} & {} & {} & {} & {} & Si-\ch{N4} & 7.6 & {} & {}\\
{} & {} & {} & {} & {} & {} & {} & Si-\ch{SiC2N} & 3.6 & {} & {}\\
{} & {} & {} & {} & {} & {} & {} & Si-\ch{C4} & 3.0 & {} & {}\\
{} & {} & {} & {} & {} & {} & {} & Other & 12.0 & {} & {}\\
\midrule
\makecell{C} & \multicolumn{2}{c}{0} & \multicolumn{2}{c}{2.1} & \multicolumn{2}{c}{42.0} & \multicolumn{2}{c}{56.0} & \multicolumn{2}{c}{0}\\
\midrule
{} & {} & {} & Other & 2.1 & C-\ch{SiC2} & 12.5 & C-\ch{Si3H} & 21.1 & {} & {}\\
{} & {} & {} & {} & {} & C-\ch{Si2C} & 10.4 & C-\ch{Si4} & 17.9 & {} & {}\\
{} & {} & {} & {} & {} & C-\ch{C3} & 6.2 & C-\ch{Si2H2} & 6.2 & {} & {}\\
{} & {} & {} & {} & {} & C-\ch{Si3} & 3.4 & C-\ch{Si2CH} & 3.4 & {} & {}\\
{} & {} & {} & {} & {} & Other & 9.5 & C-\ch{Si3C} & 3.2 & {} & {}\\
{} & {} & {} & {} & {} & {} & {} & Other & 4.2 & {} & {}\\
\midrule
\makecell{N} & \multicolumn{2}{c}{2.3} & \multicolumn{2}{c}{2.1} & \multicolumn{2}{c}{93.5} & \multicolumn{2}{c}{2.2} & \multicolumn{2}{c}{0}\\
\midrule
{} & Other & 2.3 & Other & 2.1 & N-\ch{Si3} & 71.9 & Other & 2.2 & {} & {}\\
{} & {} & {} & {} & {} & N-\ch{Si2H} & 17.0 & {} & {} & {} & {}\\
{} & {} & {} & {} & {} & N-\ch{Si2C} & 3.3 & {} & {} & {} & {}\\
{} & {} & {} & {} & {} & Other & 1.3 & {} & {} & {} & {}\\
\midrule
\makecell{H} & \multicolumn{2}{c}{100.0} & \multicolumn{2}{c}{0} & \multicolumn{2}{c}{0} & \multicolumn{2}{c}{0} & \multicolumn{2}{c}{0}\\
\midrule
{} & H-C & 63.5 & {} & {} & {} & {} & {} & {} & {} & {}\\
{} & H-N & 23.9 & {} & {} & {} & {} & {} & {} & {} & {}\\
{} & H-Si & 12.1 & {} & {} & {} & {} & {} & {} & {} & {}\\
{} & H-H & 0.5 & {} & {} & {} & {} & {} & {} & {} & {}\\
\bottomrule
\end{tabular*}
\end{table}

\begin{table}[!ht]
\small
\caption{Local atomic environments of the SiCNH8000-CL model. The "Other" sections compile environments whose individual fraction is below 3\%.}
\label{local_environments_SiCNH8000-CL}
\begin{tabular*}{\columnwidth}{@{\extracolsep\fill}lcccccccccc@{\extracolsep\fill}}
\toprule
\multicolumn{10}{c}{Proportion [\%]}\\
\makecell{Nombre de\\coordinence} & \multicolumn{2}{c}{l=1} & \multicolumn{2}{c}{l=2} & \multicolumn{2}{c}{l=3} & \multicolumn{2}{c}{l=4} & \multicolumn{2}{c}{l$\geq$5}\\
\midrule
\makecell{Si} & \multicolumn{2}{c}{0} & \multicolumn{2}{c}{0.6} & \multicolumn{2}{c}{6.8} & \multicolumn{2}{c}{88.6} & \multicolumn{2}{c}{4.1}\\
\midrule
{} & {} & {} & Other & 0.6 & Other & 6.8 & Si-\ch{CN3} & 23.3 & Other & 4.1\\
{} & {} & {} & {} & {} & {} & {} & Si-\ch{C2N2} & 20.5 & {} & {}\\
{} & {} & {} & {} & {} & {} & {} & Si-\ch{C3N} & 11.4 & {} & {}\\
{} & {} & {} & {} & {} & {} & {} & Si-\ch{N4} & 9.5 & {} & {}\\
{} & {} & {} & {} & {} & {} & {} & Si-\ch{SiCN2} & 3.5 & {} & {}\\
{} & {} & {} & {} & {} & {} & {} & Si-\ch{SiC2N} & 3.4 & {} & {}\\
{} & {} & {} & {} & {} & {} & {} & Si-\ch{C4} & 3.1 & {} & {}\\
{} & {} & {} & {} & {} & {} & {} & Si-\ch{CN2H} & 3.0 & {} & {}\\
{} & {} & {} & {} & {} & {} & {} & Other & 10.9 & {} & {}\\
\midrule
\makecell{C} & \multicolumn{2}{c}{0} & \multicolumn{2}{c}{0.5} & \multicolumn{2}{c}{47.0} & \multicolumn{2}{c}{52.5} & \multicolumn{2}{c}{0}\\
\midrule
{} & {} & {} & Other & 0.5 & C-\ch{SiC2} & 16.2 & C-\ch{Si3H} & 21.1 & {} & {}\\
{} & {} & {} & {} & {} & C-\ch{C3} & 16.1 & C-\ch{Si4} & 18.0 & {} & {}\\
{} & {} & {} & {} & {} & C-\ch{Si2C} & 4.1 & C-\ch{Si2H2} & 5.9 & {} & {}\\
{} & {} & {} & {} & {} & C-\ch{Si3} & 4.0 & Other & 7.5 & {} & {}\\
{} & {} & {} & {} & {} & C-\ch{C2H} & 3.0 & {} & {} & {} & {}\\
{} & {} & {} & {} & {} & Other & 3.6 & {} & {} & {} & {}\\
\midrule
\makecell{N} & \multicolumn{2}{c}{0.2} & \multicolumn{2}{c}{2.2} & \multicolumn{2}{c}{95.3} & \multicolumn{2}{c}{2.4} & \multicolumn{2}{c}{0}\\
\midrule
{} & Other & 0.2 & Other & 2.2 & N-\ch{Si3} & 75.6 & Other & 2.4 & {} & {}\\
{} & {} & {} & {} & {} & N-\ch{Si2H} & 16.9 & {} & {} & {} & {}\\
{} & {} & {} & {} & {} & Other & 2.8 & {} & {} & {} & {}\\
\midrule
\makecell{H} & \multicolumn{2}{c}{100.0} & \multicolumn{2}{c}{0} & \multicolumn{2}{c}{0} & \multicolumn{2}{c}{0} & \multicolumn{2}{c}{0}\\
\midrule
{} & H-C & 58.8 & {} & {} & {} & {} & {} & {} & {} & {}\\
{} & H-N & 24.4 & {} & {} & {} & {} & {} & {} & {} & {}\\
{} & H-Si & 16.4 & {} & {} & {} & {} & {} & {} & {} & {}\\
{} & H-H & 0.4 & {} & {} & {} & {} & {} & {} & {} & {}\\
\bottomrule
\end{tabular*}
\end{table}

\begin{table}[!ht]
\small
\caption{Local atomic environments of the SiCNH8000-CNB\_2200 model. The "Other" sections compile environments whose individual fraction is below 3\%.}\label{local_environments_SiCNH8000-CNBD_2200}
\begin{tabular*}{\columnwidth}{@{\extracolsep\fill}lcccccccccc@{\extracolsep\fill}}
\toprule
\multicolumn{10}{c}{Proportion [\%]}\\
\makecell{Nombre de\\coordinence} & \multicolumn{2}{c}{l=1} & \multicolumn{2}{c}{l=2} & \multicolumn{2}{c}{l=3} & \multicolumn{2}{c}{l=4} & \multicolumn{2}{c}{l$\geq$5}\\
\midrule
\makecell{Si} & \multicolumn{2}{c}{0} & \multicolumn{2}{c}{0.4} & \multicolumn{2}{c}{5.5} & \multicolumn{2}{c}{89.8} & \multicolumn{2}{c}{4.2}\\
\midrule
{} & {} & {} & Other & 0.4 & Other & 5.5 & Si-\ch{C2N2} & 27.9 & Other & 4.2\\
{} & {} & {} & {} & {} & {} & {} & Si-\ch{CN3} & 23.3 & {} & {}\\
{} & {} & {} & {} & {} & {} & {} & Si-\ch{C3N} & 14.0 & {} & {}\\
{} & {} & {} & {} & {} & {} & {} & Si-\ch{N4} & 8.3 & {} & {}\\
{} & {} & {} & {} & {} & {} & {} & Si-\ch{SiC2N} & 3.0 & {} & {}\\
{} & {} & {} & {} & {} & {} & {} & Other & 13.3 & {} & {}\\
\midrule
\makecell{C} & \multicolumn{2}{c}{0} & \multicolumn{2}{c}{0.3} & \multicolumn{2}{c}{41.7} & \multicolumn{2}{c}{58.0} & \multicolumn{2}{c}{0}\\
\midrule
{} & {} & {} & Other & 0.3 & C-\ch{C3} & 15.0 & C-\ch{Si4} & 24.2 & {} & {}\\
{} & {} & {} & {} & {} & C-\ch{SiC2} & 12.3 & C-\ch{Si3H} & 20.8 & {} & {}\\
{} & {} & {} & {} & {} & C-\ch{Si3} & 4.9 & C-\ch{Si2H2} & 7.5 & {} & {}\\
{} & {} & {} & {} & {} & C-\ch{C2H} & 3.9 & Other & 5.4 & {} & {}\\
{} & {} & {} & {} & {} & Other & 5.7 & {} & {} & {} & {}\\
\midrule
\makecell{N} & \multicolumn{2}{c}{0.2} & \multicolumn{2}{c}{2.2} & \multicolumn{2}{c}{93.9} & \multicolumn{2}{c}{3.7} & \multicolumn{2}{c}{0}\\
\midrule
{} & Other & 0.2 & Other & 2.2 & N-\ch{Si3} & 76.6 & Other & 3.7 & {} & {}\\
{} & {} & {} & {} & {} & N-\ch{Si2H} & 13.8 & {} & {} & {} & {}\\
{} & {} & {} & {} & {} & Other & 3.5 & {} & {} & {} & {}\\
\midrule
\makecell{H} & \multicolumn{2}{c}{100.0} & \multicolumn{2}{c}{0} & \multicolumn{2}{c}{0} & \multicolumn{2}{c}{0} & \multicolumn{2}{c}{0}\\
\midrule
{} & H-C & 65.3 & {} & {} & {} & {} & {} & {} & {} & {}\\
{} & H-N & 21.6 & {} & {} & {} & {} & {} & {} & {} & {}\\
{} & H-Si & 11.9 & {} & {} & {} & {} & {} & {} & {} & {}\\
{} & H-H & 1.2 & {} & {} & {} & {} & {} & {} & {} & {}\\
\bottomrule
\end{tabular*}
\end{table}

\clearpage
\newpage

\section{Visualization of the progressive evolution of carbon domain organization}

\begin{figure}[!ht]
\centering
\includegraphics[width=0.99\linewidth]{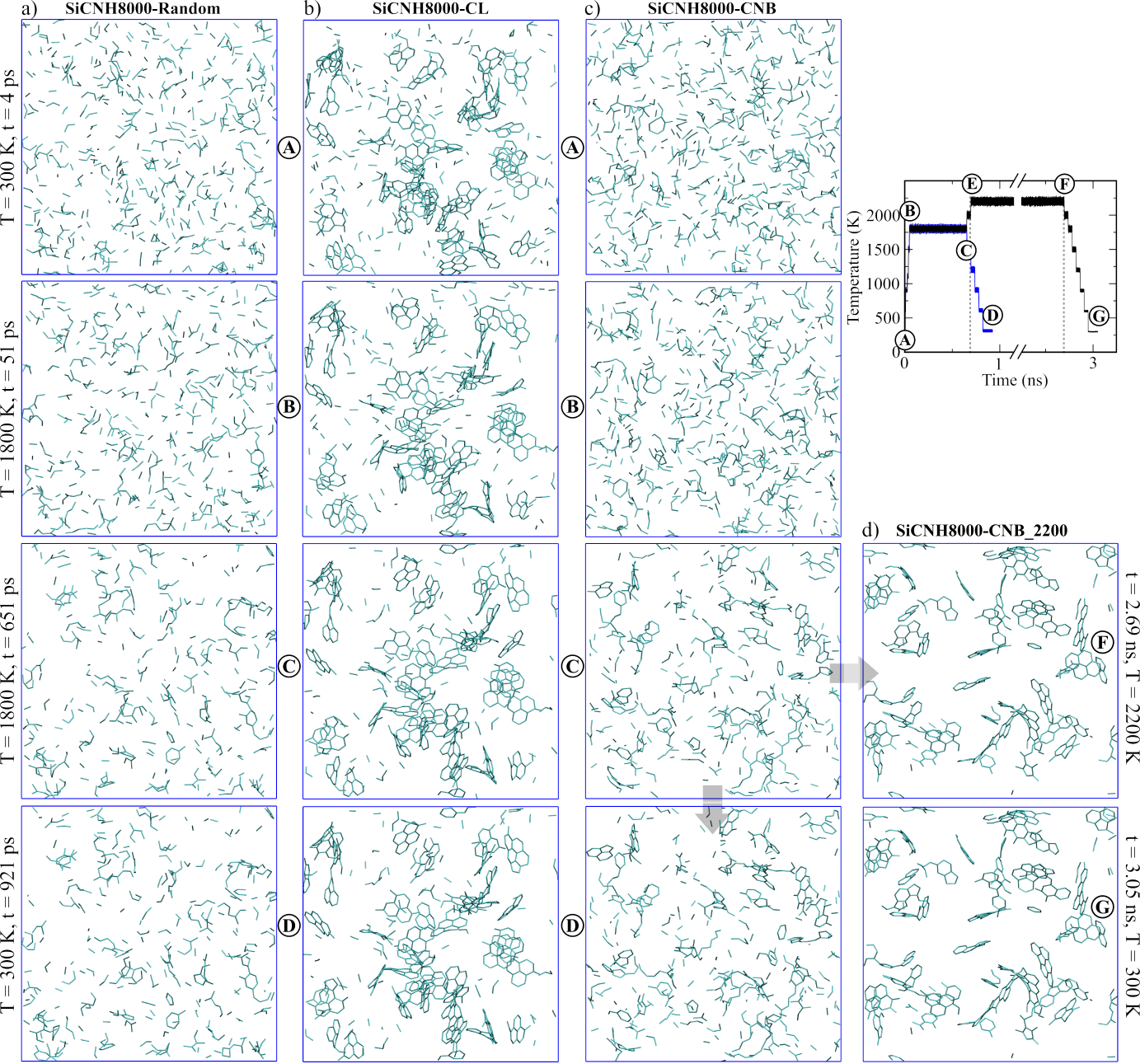}
\caption{Configuration of C-C bonds in the (a) SiCNH8000-Random, (b)  SiCNH8000-CL, (c) SiCNH8000-CNB, and (d) SiCNH8000-CNB\_2200 models at different stages (A-D for the former three models and F-G for the latter, heated up to 2200 K) of the thermal cycles, shown for convenience on the top right side of the figure. The progressive evolution of carbon domain organization is clearly visible, with significant differences in the initial configurations leading to distinct pathways of carbon structuration.}
\label{figS:snapshots_8000}
\end{figure}

\clearpage
\newpage

\section{Evolution of the number of carbon rings}

\begin{figure}[!ht]
\centering
\includegraphics[width=0.45\linewidth]{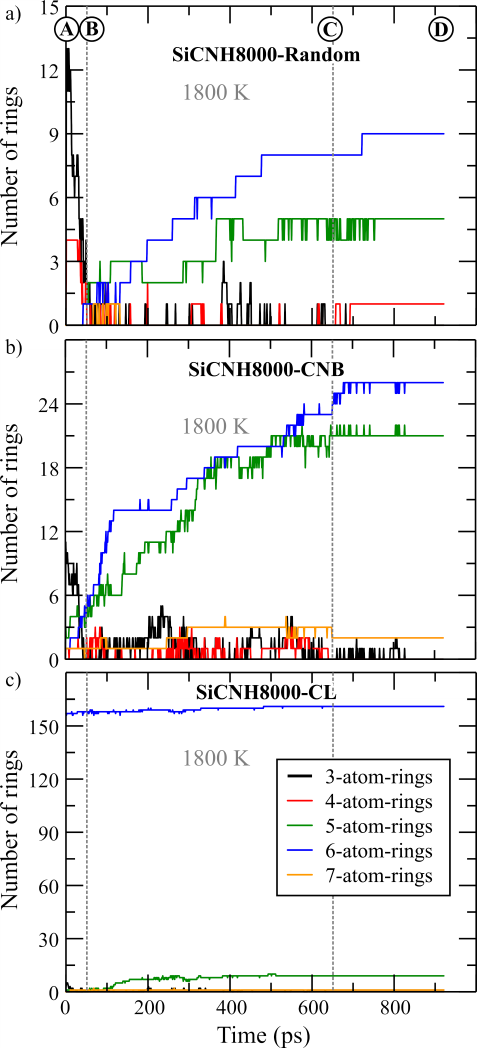}
\caption{Evolution of the number of carbon rings with different sizes during the thermal cycle. The vertical dashed lines indicate the 1800 K temperature plateau, marking a period of accelerated ring formation and transformation. Circled A-D capital letters indicate the timing of configurations shown in Figure S\ref{fig:snapshots_8000}.}
\label{fig:ring_evolution}
\end{figure}


\end{document}